\pdfoutput=1
\RequirePackage{ifpdf}
\ifpdf 
\documentclass[pdftex]{sigma}
\else
\documentclass{sigma}
\fi

\numberwithin{equation}{section}

\begin{document}

\newcommand{\arXivNumber}{1301.2401}

\allowdisplaybreaks

\renewcommand{\PaperNumber}{090}

\FirstPageHeading

\ShortArticleName{Hypergeometric Solutions of the $A_4^{(1)}$-Surface~$q$-Painlev\'e IV Equation}

\ArticleName{Hypergeometric Solutions\\
of the $\boldsymbol{A_4^{(1)}}$-Surface $\boldsymbol{q}$-Painlev\'e IV Equation}

\Author{Nobutaka NAKAZONO}

\AuthorNameForHeading{N.~Nakazono}

\Address{School of Mathematics and Statistics, The University of Sydney,\\
New South Wales 2006, Australia}
\Email{\href{mailto:nobua.n1222@gmail.com}{nobua.n1222@gmail.com}}

\ArticleDates{Received June 06, 2013, in f\/inal form August 14, 2014; Published online August 22, 2014}

\Abstract{We consider a~$q$-Painlev\'e IV equation which is the $A_4^{(1)}$-surface type in the Sakai's classif\/ication.
We f\/ind three distinct types of classical solutions with determinantal structures whose elements are basic
hypergeometric functions.
Two of them are expressed by~${}_2\varphi_1$ basic hypergeometric series and the other is given by ${}_2\psi_2$
bilateral basic hypergeometric series.}


\Classification{33D05; 33D15; 33D45; 33E17; 39A13}

\section{Introduction}

The focus of this paper is on the following single second-order ordinary dif\/ference equation:
\begin{gather}
 (X_{n+1}X_{n}-1)(X_{n-1}X_n-1)
\nonumber
\\
\qquad
 =q^{(-N+2n-m-1)/2}a_0a_1^{3/2}a_2^2 \frac{\big(X_n+q^{(N-m)/2}a_1^{1/2}\big)\big(X_n+q^{(-N+m)/2}a_1^{-1/2}\big)}
{X_n+q^{(-N+n-m)/2}a_1^{1/2}a_2},
\label{eqn:qp4}
\end{gather}
where $n\in\mathbb{Z}$ is the independent variable, $X_n=X_n(m,N)$ is the dependent variable and $m,N\in\mathbb{Z}$ and
$a_0$, $a_1$, $a_2$, $q$ $(|q|<1)\in\mathbb{C}^\ast$ are parameters.
Equation~\eqref{eqn:qp4} is known as a~$q$-discrete analog of the Painlev\'e IV equation ($q$-P$_{\rm IV}$)~\cite{RG1996:MR1399286}.

In 2001, Sakai introduced a~geometric approach to the theory of the Painlev\'e and discrete Painlev\'e equations
(Painlev\'e systems) and showed the classif\/ications of Painlev\'e systems by the rational
surface~\cite{SakaiH2001:MR1882403}.
The rational surface can be identif\/ied with the space of initial condition, and the group of Cremona isometries
associated with the surface generate the af\/f\/ine Weyl group.
He also showed that the translation part of the af\/f\/ine Weyl group gives rise to various discrete Painlev\'e equations.
Then, such discrete Painlev\'e equations are said to have the af\/f\/ine Weyl group symmetries.

In 2004,~$q$-P$_{\rm IV}$~\eqref{eqn:qp4} was generalized to the following simultaneous f\/irst-order ordinary dif\/fe\-rence
equations by the singularity conf\/inement criterion~\cite{TGCR2004:MR2058894}:
\begin{subequations}
\label{eqn:qp5}
\begin{gather}
 (y_{k+1}x_k-1)(y_kx_k-1)
\nonumber
\\
\quad
 =q^{(-N+4k-m+2l-2)/2}a_0a_1^{3/2}a_2^2a_3 \frac{\big(x_k+q^{(N-m)/2}a_1^{1/2}\big)\big(x_k+q^{(-N+m)/2}a_1^{-1/2}\big)}
{x_k+q^{(-N+2k-m+2l)/2}a_1^{1/2}a_2},
\\
 (y_kx_k-1)(y_kx_{k-1}-1)
\nonumber
\\
\quad
 =q^{(-N+4k-m+2l-4)/2}a_0a_1^{3/2}a_2^2a_3 \frac{\big(y_k+q^{(N-m)/2}a_1^{1/2}\big)\big(y_k+q^{(-N+m)/2}a_1^{-1/2}\big)}
{y_k+q^{(-N+2k-m-2)/2}a_1^{1/2}a_2a_3},
\end{gather}
\end{subequations}
where \looseness=-1 $k\in\mathbb{Z}$ is the independent variable, $x_k=x_k(l,m,N)$ and $y_k=y_k(l,m,N)$ are dependent variables and
$l,m,N\in\mathbb{Z}$ and $a_0$, $a_1$, $a_2$, $a_3$, $q$ $(|q|<1)\in\mathbb{C}^\ast$ are parameters.
System~\eqref{eqn:qp5} is known as a~$q$-discrete analog of the Painlev\'e V equation ($q$-P$_{\rm V}$).
It is also known that~$q$-P$_{\rm V}$~\eqref{eqn:qp5} is the $A_4^{(1)}$-surface type in the Sakai's classif\/ication and
has the af\/f\/ine Weyl group symmetry of type~$A_4^{(1)}$.

Conversely,~$q$-P$_{\rm IV}$~\eqref{eqn:qp4} can be recovered from~$q$-P$_{\rm V}$~\eqref{eqn:qp5} by putting
\begin{gather*}
a_3=q^{1/2},
\qquad
l=0,
\end{gather*}
and replacing the independent variable and the dependent variables~by
\begin{gather*}
2k=n,
\qquad
x_k=X_n,
\qquad
y_k=X_{n-1}.
\end{gather*}
This procedure is referred to as ``{\it symmetrization}'' of~$q$-P$_{\rm V}$~\eqref{eqn:qp5}, which comes from the
terminology of the Quispel--Roberts--Thompson (QRT) mapping~\cite{QRT1988:MR924318,QRT1989:MR982386}.
After this terminology, \mbox{$q$-P$_{\rm V}$}~\eqref{eqn:qp5} is sometimes called the ``{\it asymmetric discrete Painlev\'e
equation}'', and \mbox{$q$-P$_{\rm IV}$}~\eqref{eqn:qp4} is called the ``{\it symmetric discrete Painlev\'e equation}''.
It appears as though the symmetrization is a~simple specialization on the level of the equation, but the following
problems were known:
\begin{enumerate}\itemsep=0pt
\item[(i)] According \looseness=-1 to Sakai's theory~\cite{SakaiH2001:MR1882403}, the discrete Painlev\'e equations arise as the
birational mappings corresponding to the translations of the af\/f\/ine Weyl groups.
The asymmetric discrete Painlev\'e equations are characterized in this manner, however, it was not known how to
characterize the symmetric discrete Painlev\'e equations as the action of af\/f\/ine Weyl groups;
\item[(ii)] Painlev\'e systems admit the particular solutions expressible in terms of the hypergeometric type functions
({\it hypergeometric solutions}) when some of the parameters take special values (see, for
example,~\cite{KMNOY2004:MR2077840, KMNOY2005:MR2153786} and references therein).
However, the hypergeometric solutions to the symmetric discrete Painlev\'e equation cannot be obtained by the na\"ive
specialization of those to the corresponding asymmetric equation.
\end{enumerate}

In~\cite{KNT2011:MR2773334}, the mechanism of the symmetrization was investigated and the nontrivial inconsistency among
the hypergeometric solutions were explained in detail by taking an example of~$q$-Painlev\'e equation with the af\/f\/ine
Weyl group symmetry of type $(A_2+A_1)^{(1)}$.
The key to characterize the symmetric discrete Painlev\'e equation as the action of af\/f\/ine Weyl group is taking the
half-step translation instead of a~translation as a~time evolution.
In general, various discrete dynamical systems of Painlev\'e type can be obtained from elements of inf\/inite order that
are not necessarily translations in the af\/f\/ine Weyl group by taking the projection on appropriate subspaces of the
parameter spaces.
Such a~procedure is called a~``{\it projective reduction}''.

It is well known that the~$\tau$-functions play a~crucial role in the theory of integrable
systems~\cite{book_MJD2000:MR1736222}, and it is also possible to introduce them in the theory of Painlev\'e
systems~\cite{JM1981:MR625446,JM1981:MR636469,JMU1981:MR630674,book_NoumiM2004:MR2044201,OkamotoK1986:MR854008,
OkamotoK1987:MR916698,OkamotoK1987:MR914314,OkamotoK1987:MR927186}.
A~representation of the af\/f\/ine Weyl groups can be lifted on the level of
the~$\tau$-functions~\cite{KMNOY2003:MR1984002,KMNOY2006:MR2353465, TsudaT2006:MR2207047}, which gives rise to various
bilinear equations of Hirota type satisf\/ied the~$\tau$-functions.
Usually, the hypergeometric solutions are derived by reducing the bilinear equations to the Pl\"ucker relations by using
the contiguity relations satisf\/ied by the entries of
determinants~\cite{HK2007:MR2392886,HKW2006:MR2264726,JKM2006:MR2297948,KK2003:MR1952868,KM1999:MR1694666,KNY2001:MR1876614,KO1998:MR1629434,
KOS1995:MR1341981,KOSGR1994:MR1267458,ON1992:MR1200649,
SakaiH1998:MR1632613}.
This method is elementary, but it encounters technical dif\/f\/iculties for Painlev\'e systems with large symmetries.
In order to overcome this dif\/f\/iculty, Masuda has proposed a~method of constructing hypergeometric solutions under
a~certain boundary condition on the lattice where the~$\tau$-functions live ({\it hypergeometric~$\tau$-functions}), so
that they are consistent with the action of the af\/f\/ine Weyl
groups~\cite{MasudaT2009:MR2506177,MasudaT2011:MR2765599,NakazonoN2010:MR2769931}.
Although this requires somewhat complex calculations, the merit is that it is systematic and that it can be applied to
the systems with large symmetries.

In~\cite{KN:inpress}, the list of the simplest hypergeometric solutions to the symmetric~$q$-Painlev\'e equations are
shown.
In general, hypergeometric solutions of Painlev\'e systems can be expressed by determinants whose entries are given by
hypergeometric type functions.
Therefore, it is natural to be curious about the determinant formulae of them.
The purpose of this paper is to obtain the determinant formulae of the hypergeometric solutions to the~$q$-P$_{\rm IV}$
via the construction of the hypergeometric~$\tau$-functions and the theory of orthogonal polynomials.

This paper is organized as follows: in Section~\ref{section:weyl_group_A4}, we f\/irst introduce a~representation of the
af\/f\/ine Weyl group of type $A_4^{(1)}$.
Next, we show how~$q$-P$_{\rm V}$~\eqref{eqn:qp5} and~$q$-P$_{\rm IV}$~\eqref{eqn:qp4} can be derived from the
representation.
In Section~\ref{section:hyper_sol_qP4_1}, we construct the hypergeometric~$\tau$-functions for the~$q$-P$_{\rm IV}$ and
obtain the hypergeometric solutions of the~$q$-P$_{\rm IV}$ which are expressed by basic hypergeometric series (see
Theorems~\ref{theorem:lattice_sol_1} and~\ref{theorem:lattice_sol_2}).
In Section~\ref{section:hyper_sol_qP4_2}, we obtain the hypergeometric solutions of the~$q$-P$_{\rm IV}$ which are
expressed by bilateral basic hypergeometric series via the theory of orthogonal polynomials (see
Theorem~\ref{theorem:molecule_sol}).
Some concluding remarks are given in Section~\ref{section:concluding}.

We use the following conventions of~$q$-analysis with $|q|,|p|<1$ throughout this
paper~\cite{book_GR2004:MR2128719,book_KLS2010:MR2656096}:
\begin{itemize}\itemsep=0pt
\item~$q$-shifted factorials:
\begin{gather*}
(a;q)_\infty=\prod\limits_{i=1}^\infty\big(1-aq^{i-1}\big),
\qquad
(a;q)_\lambda=\frac{(a;q)_\infty}{(aq^\lambda;q)_\infty},
\end{gather*}
where $\lambda\in\mathbb{C}$;
\item Jacobi theta function:
\begin{gather*}
\Theta(a;q)=(a;q)_\infty\big(qa^{-1};q\big)_\infty;
\end{gather*}
\item Elliptic gamma function:
\begin{gather*}
\Gamma(a;p,q)=\frac{\big(pqa^{-1};p,q\big)_{\infty}}{(a;p,q)_{\infty}},
\end{gather*}
where
\begin{gather*}
(a;p,q)_k=\prod\limits_{i,j=0}^{k-1} \big(1-p^iq^ja\big);
\end{gather*}
\item Basic hypergeometric series:
\begin{gather*}
{}_s\varphi_r\left(
\begin{matrix}
a_1,\dots,a_s
\\
b_1,\dots,b_r
\end{matrix}
;q,z\right) =\sum\limits_{n=0}^{\infty}\frac{(a_1,\dots,a_s;q)_n}{(b_1,\dots,b_r;q)_n(q;q)_n}
\big[
(-1)^nq^{n(n-1)/2}
\big]
^{1+r-s}z^n,
\end{gather*}
where
\begin{gather*}
(a_1,\dots,a_s;q)_n=\prod\limits_{j=1}^{s}(a_j;q)_n;
\end{gather*}
\item Bilateral basic hypergeometric series:
\begin{gather*}
{}_s\psi_r\left(
\begin{matrix}
a_1,\dots,a_s
\\
b_1,\dots,b_r
\end{matrix}
;q,z\right) =\sum\limits_{n=-\infty}^{\infty}\frac{(a_1,\dots,a_s;q)_n}{(b_1,\dots,b_r;q)_n}
\big[
(-1)^nq^{n(n-1)/2}
\big]
^{r-s}z^n;
\end{gather*}
\item Bilateral~$q$-integral:
\begin{gather*}
\int^\infty_{-\infty}f(t){\rm d}_qt =(1-q)\sum\limits_{n=-\infty}^\infty \big(f(q^n)+f(-q^n)\big)q^n.
\end{gather*}
\end{itemize}
We note that the following formulae hold:
\begin{gather*}
\frac{(a;q)_{\lambda+1}}{(a;q)_\lambda}=1-aq^\lambda,
\qquad
\frac{\Theta(qa;q)}{\Theta(a;q)}=-a^{-1},
\qquad
\frac{\Gamma(qa;q,q)}{\Gamma(a;q,q)}=\Theta(a;q).
\end{gather*}

\section[Af\/f\/ine Weyl group of type $A_4^{(1)}$]{Af\/f\/ine Weyl group of type $\boldsymbol{A_4^{(1)}}$}
\label{section:weyl_group_A4}

\subsection[Birational representation of the af\/f\/ine Weyl group of type $A_4^{(1)}$]{Birational representation
of the af\/f\/ine Weyl group of type $\boldsymbol{A_4^{(1)}}$}

In this section, we formulate the family of B\"acklund transformations of~$q$-P$_{\rm V}$~\eqref{eqn:qp5} as
a~birational representation of the af\/f\/ine Weyl group of type $A_4^{(1)}$.

Let $s_i$ $(i=0,1,2,3,4)$,~$\sigma$ and~$\iota$ be transformations of the parameters $a_k$ $(k=0,1,2,3,4)$ and the
variables $f_j$ $(j=0,1,2,3,4)$.
The action of the transformations on the parameters is given~by
\begin{gather*}
s_i(a_j)=a_ja_i^{-a_{ij}},
\qquad
\sigma(a_i)=a_{i+1},
\\
\iota: \ (a_0,a_1,a_2,a_3,a_4)\mapsto \big(a_0^{-1},a_4^{-1},a_3^{-1},a_2^{-1},a_1^{-1}\big),
\end{gather*}
where $i,j\in\mathbb{Z}/5\mathbb{Z}$ and the symmetric $5\times 5$ matrix
\begin{gather*}
A=(a_{ij})_{i,j=0}^4 =\left(
\begin{matrix}
2&-1&0&0&-1
\\
-1&2&-1&0&0
\\
0&-1&2&-1&0
\\
0&0&-1&2&-1
\\
-1&0&0&-1&2
\end{matrix}
\right)
\end{gather*}
is the Cartan matrix of type $A_4^{(1)}$.
Moreover, the action on the variables is given~by
\begin{gather*}
 s_i(f_{i+2})=\frac{a_{i+3}a_{i+4}(a_ia_{i+1}+a_{i+3}f_i)}{a_{i+1}^2f_{i+3}},
\qquad
s_i(f_{i+4})=\frac{a_{i+4}(a_{i+2}+a_{i+4}a_if_{i+1})}{a_ia_{i+1}a_{i+2}^2f_{i+3}},
\\
 s_i(f_j)=f_j,
\quad
j\neq i+2,i+4,
\qquad
\sigma(f_i)=f_{i+1},
\\
 \iota: \ (f_0,f_1,f_2,f_3,f_4)\mapsto (f_1,f_0,f_4,f_3,f_2),
\end{gather*}
where $i\in\mathbb{Z}/5\mathbb{Z}$.
Note that the variables satisfy the following conditions:
\begin{gather*}
a_{i+3}^2a_{i+4}f_i=a_{i+1}(a_ia_{i+1}f_{i+2}f_{i+3}-a_{i+3}a_{i+4}),
\end{gather*}
where $i\in\mathbb{Z}/5\mathbb{Z}$.
The conditions above look like f\/ive, but they are essentially three.
Therefore, variables $f_i$ are essentially two.
\begin{proposition}[\cite{HK2007:MR2392886,SakaiH2001:MR1882403,TsudaT2006:MR2207047}]
The group of birational transformations $\widetilde{W}\big(A_4^{(1)}\big)=\langle s_0,s_1,s_2,s_3,$ $s_4,\sigma,\iota\rangle$
gives a~representation of the $($extended$)$ affine Weyl group of type~$A_4^{(1)}$.
Namely, the transformations satisfy the fundamental relations
\begin{gather*}
 s_i^2=1,
\qquad
(s_is_{i\pm 1})^3=1,
\qquad
(s_is_j)^2=1,
\quad
j\neq i\pm 1,
\qquad
\sigma^5=1,
\qquad
\sigma s_i=s_{i+1}\sigma,
\\
 \iota^2=1,
\qquad
\iota s_0=s_0\iota,
\qquad
\iota s_1=s_4\iota,
\qquad
\iota s_2=s_3\iota,
\end{gather*}
where $i,j\in\mathbb{Z}/5\mathbb{Z}$.
\end{proposition}

In general, for a~function $F=F(a_i,f_j)$, we let an element $w\in\widetilde{W}\big(A_4^{(1)}\big)$ act as
$w.F(a_i,f_j)=F(w.a_i,w.f_j)$, that is,~$w$ acts on the arguments from the left.
Note that $q=a_0a_1a_2a_3a_4$ is invariant under the action of $\langle s_0,s_1,s_2,s_3,s_4,\sigma\rangle$.
We def\/ine the translations $T_i$ ($i=0,1,2,3,4$)~by
\begin{gather}
\label{eqn:translations}
T_0=\sigma s_4s_3s_2s_1,
\qquad
T_1=\sigma s_0s_4s_3s_2,
\qquad
T_2=\sigma s_1s_0s_4s_3,
\nonumber
\\
T_3=\sigma s_2s_1s_0s_4,
\qquad
T_4=\sigma s_3s_2s_1s_0,
\end{gather}
whose action on the parameters is given~by
\begin{gather*}
\begin{matrix}
T_0: \ (a_0,a_1,a_2,a_3,a_4)\mapsto\big(qa_0,q^{-1}a_1,a_2,a_3,a_4\big),
\\
T_1: \ (a_0,a_1,a_2,a_3,a_4)\mapsto\big(a_0,qa_1,q^{-1}a_2,a_3,a_4\big),
\\
T_2: \ (a_0,a_1,a_2,a_3,a_4)\mapsto\big(a_0,a_1,qa_2,q^{-1}a_3,a_4\big),
\\
T_3: \ (a_0,a_1,a_2,a_3,a_4)\mapsto\big(a_0,a_1,a_2,qa_3,q^{-1}a_4\big),
\\
T_4: \ (a_0,a_1,a_2,a_3,a_4)\mapsto\big(q^{-1}a_0,a_1,a_2,a_3,qa_4\big).
\end{matrix}
\end{gather*}
Note that $T_i$ ($i=0,1,2,3,4$) commute with each other and $T_0T_1T_2T_3T_4=1$.

\subsection[Derivations of the~$q$-Painlev\'e equations]{Derivations of the~$\boldsymbol{q}$-Painlev\'e equations}

In this section, we derive the~$q$-Painlev\'e equations from $\widetilde{W}\big(A_4^{(1)}\big)$.
The action of $T_{23}=T_2T_3$ on $f$-variables can be expressed as
\begin{gather}
 (T_{23}(y)x-1)(yx-1) =q^{-1}a_0a_1^{3/2}a_2^2a_3 \frac{\big(x+a_1^{1/2}\big)\big(x+a_1^{-1/2}\big)}{x+a_1^{1/2}a_2},
\label{eqn:action_T23_1}
\\
 (yx-1)\big(yT_{23}^{-1}(x)-1\big) =q^{-2}a_0a_1^{3/2}a_2^2a_3
\frac{\big(y+a_1^{1/2}\big)\big(y+a_1^{-1/2}\big)}{y+q^{-1}a_1^{1/2}a_2a_3},
\label{eqn:action_T23_2}
\end{gather}
where
\begin{gather*}
x=a_0a_1^{1/2}a_3^{-1}f_2,
\qquad
y=a_1^{-1/2}a_2^{-1}a_4^{-1}s_4(f_1).
\end{gather*}
Applying $T_{23}^kT_2^lT_0^mT_1^N$ on equations~\eqref{eqn:action_T23_1} and~\eqref{eqn:action_T23_2} and
putting
\begin{gather*}
x_k(l,m,N)=T_{23}^kT_2^lT_0^mT_1^N(x),
\qquad
y_k(l,m,N)=T_{23}^kT_2^lT_0^mT_1^N(y),
\end{gather*}
we obtain~$q$-P$_{\rm V}$~\eqref{eqn:qp5}.
Then, we can regard $T_{23}$ and $T_i$ ($i=0,1,2,3,4$) as the time evolution and the B\"acklund transformations
of~$q$-P$_{\rm V}$~\eqref{eqn:qp5}, respectively.
We note that considering the action of $T_0$:
\begin{gather*}
 T_0(g)g =\frac{\big(f+a_0^{-3/4}a_1^{1/4}a_3^{-1/4}\big)\big(f+a_0^{-3/4}a_1^{1/4}a_3^{-1/4}a_4^{-1}\big)}
{1+a_0^{-1/4}a_1^{-1/4}a_3^{1/4}f},
\\
 T_0^{-1}(f)f =\frac{\big(g+a_0^{-1/4}a_1^{3/4}a_3^{1/4}\big)\big(g+a_0^{-1/4}a_1^{3/4}a_2a_3^{1/4}\big)}
{1+a_0^{1/4}a_1^{1/4}a_3^{-1/4}g},
\end{gather*}
where
\begin{gather*}
f=a_0^{-3/4}a_1^{-3/4}a_3^{3/4}f_0,
\qquad
g=a_0^{3/4}a_1^{3/4}a_3^{-3/4}f_2,
\end{gather*}
we obtain another~$q$-discrete analog of Painlev\'e V equation~\cite{SakaiH2001:MR1882403}:
\begin{subequations}
\label{eqn:qp5_sakai}
\begin{gather}
 g_{n{+}1}g_n  =  \frac{\big(f_n\!+\!q^{{-}n{+}\frac{k{+}l{-}m}{4}}a_0^{{-}\frac{3}{4}}a_1^{\frac{1}{4}}a_3^{{-}\frac{1}{4}}\big)
\big(f_n\!+\!q^{{-}n{+}\frac{k{+}l{+}3m}{4}}a_0^{{-}\frac{3}{4}}a_1^{\frac{1}{4}}a_3^{{-}\frac{1}{4}}a_4^{{-}1}\big)}
{1\!+\!q^{\frac{{-}k{-}l{+}m}{4}}a_0^{{-}\frac{1}{4}}a_1^{{-}\frac{1}{4}}a_3^{\frac{1}{4}}f_n},
\\
 f_{n{+}1}f_n =\frac{\big(g_{n{+}1}\!+\!q^{{-}n{-}1{+}\frac{3k{-}l{+}m}{4}}a_0^{{-}\frac{1}{4}}a_1^{\frac{3}{4}}a_3^{\frac{1}{4}}\big)
 \big(g_{n+1}+q^{{-}n{-}1{+}\frac{{-}k{+}3l{+}m}{4}}a_0^{{-}\frac{1}{4}}a_1^{\frac{3}{4}}a_2a_3^{\frac{1}{4}}\big)}
{1\!+\!q^{\frac{k{+}l{-}m}{4}}a_0^{\frac{1}{4}}a_1^{\frac{1}{4}}a_3^{{-}\frac{1}{4}}g_{n+1}},
\end{gather}
\end{subequations}
where{\samepage
\begin{gather*}
f_n=f_n(k,l,m)=T_0^nT_1^kT_2^l T_3^m(f),
\qquad
g_n=g_n(k,l,m)=T_0^nT_1^kT_2^l T_3^m(g).
\end{gather*}
Thus,~$q$-P$_{\rm V}$~\eqref{eqn:qp5} and equation~\eqref{eqn:qp5_sakai} are the B\"acklund transformations each other.}

In order to derive~$q$-P$_{\rm IV}$~\eqref{eqn:qp4}, we factorize $T_{23}$ as follows
\begin{gather*}
T_{23}=R_{23}^2,
\end{gather*}
where $R_{23}$ is given~by
\begin{gather}
\label{eqn:R23}
R_{23}=\sigma s_1s_0s_4.
\end{gather}
The action of $R_{23}$ on the parameters is given~by
\begin{gather*}
R_{23}: \ (a_0,a_1,a_2,a_3,a_4)\mapsto\big(a_0,a_1,a_2a_3,qa_3^{-1},q^{-1}a_3a_4\big).
\end{gather*}
Let us consider the projection of the action of $R_{23}$ on the line
\begin{gather}
\label{eqn:condition_R23}
a_3=q^{1/2}.
\end{gather}
Then, the action on the parameters becomes translational motion:
\begin{gather*}
R_{23}: \ (a_0,a_1,a_2,a_4)\mapsto\big(a_0,a_1,q^{1/2}a_2,q^{-1/2}a_4\big).
\end{gather*}
Since the action of $R_{23}$ on the variable $f_2$ is given~by
\begin{gather*}
 R_{23}(f_2)
=\frac{q}{a_0^{2}a_1f_2}\left(1+\frac{a_0\big(a_0f_2+q^{1/2}\big)\big(a_0a_1f_2+q^{1/2}\big)}{q^{1/2}\big(q^{1/2}a_2+a_0f_2\big)f_4}\right),
\\
 R_{23}^{-1}(f_2)=\frac{q}{a_0^2a_1f_2}\left(1+\frac{a_1a_2^2}{q^{1/2}}f_4\right),
\end{gather*}
we obtain
\begin{gather}
\label{eqn:action_R23_3}
\big(R_{23}(X)X-1\big)\big(R_{23}^{-1}(X)X-1\big)
=q^{-1/2}a_0a_1^{3/2}a_2^2\frac{\big(X+a_1^{1/2}\big)\big(X+a_1^{-1/2}\big)}{X+a_1^{1/2}a_2},
\end{gather}
where
\begin{gather}
\label{eqn:relation_X_f2}
X=q^{-1/2}a_0a_1^{1/2}f_2.
\end{gather}
Applying $R_{23}^nT_0^mT_1^N$ on equation~\eqref{eqn:action_R23_3} and putting
\begin{gather*}
X_n(m,N)=R_{23}^nT_0^mT_1^N(X),
\end{gather*}
we obtain~$q$-P$_{\rm IV}$~\eqref{eqn:qp4}.
Note that $R_{23}$ commute with $T_i$ $(i=0,1,4)$ and $T_0T_1R_{23}^2T_4=1$.
Then, $R_{23}$ and $T_i$ $(i=0,1,4)$ are regarded as the time evolution and the B\"acklund transformations of~$q$-P$_{\rm IV}$~\eqref{eqn:qp4},
respectively.

\section[Hypergeometric solutions of the~$q$-P$_{\rm IV}$ (I)]{Hypergeometric solutions
of the~$\boldsymbol{q}$-P$\boldsymbol{_{\rm IV}}$ (I)}
\label{section:hyper_sol_qP4_1}

In this section, we obtain the hypergeometric solutions of~$q$-P$_{\rm IV}$~\eqref{eqn:qp4} by constructing the
hypergeometric~$\tau$-functions for the~$q$-P$_{\rm IV}$.

\subsection[$\tau$-functions]{$\boldsymbol{\tau}$-functions}

In this section, we def\/ine the~$\tau$-functions.

We introduce the new variables $\tau_i$ $(i=1,2,\dots,7)$ with
\begin{gather}
\label{eqn:relation_f2_tau}
f_2=\frac{\tau_4\tau_5}{\tau_6\tau_7},
\qquad
f_4=\frac{\tau_1\tau_2}{\tau_3\tau_7},
\end{gather}
and lift the representation of $\widetilde{W}\big(A_4^{(1)}\big)$ on their level:
\begin{gather*}
 s_0(\tau_1)=\frac{a_4 (a_0\tau_3\tau_4\tau_5+a_2a_3\tau_1\tau_2
\tau_6+a_0a_3\tau_3\tau_6\tau_7 )}{a_0^2a_1a_2\tau_4\tau_7},
\qquad
s_0(\tau_i)=\tau_i, \quad i=2,3,5,6,
\\
 s_0(\tau_4)=\frac{a_0a_4 (a_0\tau_3\tau_4\tau_5+a_2a_3\tau_1\tau_2\tau_6
+a_3\tau_3\tau_6\tau_7 )}{a_1a_2\tau_1\tau_7},
\\
 s_0(\tau_7)=\frac{a_4\big(a_0^2\tau_3\tau_4\tau_5+a_3a_0\tau_3\tau_6\tau_7
+a_2a_3\tau_1\tau_2\tau_6\big)}{a_0a_1a_2\tau_1\tau_4},
\qquad
s_1(\tau_1)=\tau_2,
\qquad
s_1(\tau_2)=\tau_1,
\\
 s_1(\tau_i)=\tau_i, \quad i=3,\dots,7,
\qquad
s_2(\tau_1)=\frac{a_0a_1 (a_0\tau_4\tau_5+a_2a_3\tau_6\tau_7 )}{a_3^2\tau_3},
\\
 s_2(\tau_3)=\frac{a_0a_1 (a_0\tau_4\tau_5+a_3\tau_6\tau_7 )}{a_2a_3^2\tau_1},
\qquad
s_2(\tau_i)=\tau_i, \quad i=2,4,5,6,7,
\\
 s_3(\tau_4)=\frac{a_2 (a_2a_3\tau_1\tau_2+a_0\tau_3\tau_7 )} {a_0^2a_3a_4\tau_6},
\qquad
s_3(\tau_6)=\frac{a_2a_3 (a_2\tau_1\tau_2+a_0\tau_3\tau_7 )}{a_0^2a_4\tau_4},
\\
 s_3(\tau_i)=\tau_i, \quad i=1,2,3,5,7,
\qquad
s_4(\tau_4)=s_4(\tau_5),
\qquad
s_4(\tau_5)=s_4(\tau_4),
\\
 s_4(\tau_i)=\tau_i, \quad i=1,2,3,6,7,
\qquad
\iota: \ (\tau_1,\tau_2,\tau_3,\tau_4,\tau_5,\tau_6,\tau_7) =(\tau_4,\tau_5,\tau_6,\tau_1,\tau_2,\tau_3,\tau_7),
\\
 \sigma(\tau_1)=\frac{a_0a_1\left(a_0\tau_4\tau_5+a_3\tau_6\tau_7\right)} {a_2a_3^2\tau_1},
\qquad
\sigma(\tau_2)=\tau_3,
\qquad
\sigma(\tau_3)=\tau_6,
\\
 \sigma(\tau_4)=\frac{a_4 \big(a_0^2\tau_3\tau_4\tau_5+a_3a_0\tau_3\tau_6\tau_7
+a_2a_3\tau_1\tau_2\tau_6 \big)}{a_0a_1a_2\tau_1\tau_4},
\qquad
\sigma(\tau_5)=\tau_7,
\\
 \sigma(\tau_6)=\tau_5,
\qquad
\sigma(\tau_7)=\tau_2.
\end{gather*}
Then, we get the following proposition:
\begin{proposition}
[\cite{TsudaT2006:MR2207047}]
\label{prop:weyl_group_tau}
The transformations: $s_0$, $s_1$, $s_2$, $s_3$, $s_4$, $\sigma$, $\iota$, on the variables $\tau_i$ $(i=1,2,\dots,7)$ also realize the
$($extended$)$ affine Weyl group of type $A_4^{(1)}$.
\end{proposition}

\begin{figure}[h]
\centering
\includegraphics[width=0.65\textwidth]{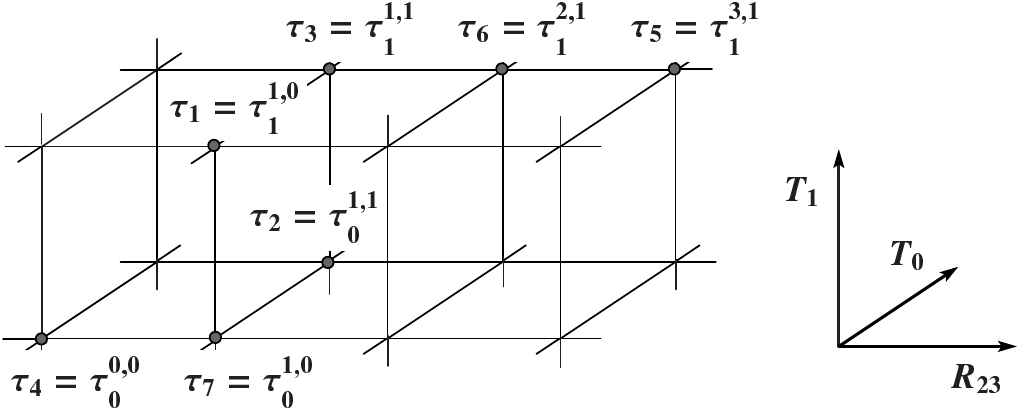}
\caption{Conf\/iguration of the~$\tau$-functions on the 3D-lattice.}
\label{fig:configuration_tau}
\end{figure}

Let us def\/ine the~$\tau$-functions $\tau^{n,m}_{N}$ ($n,m,N\in\mathbb{Z}$)~by
\begin{gather*}
\tau_N^{n,m}=R_{23}^nT_0^mT_1^N(\tau_4).
\end{gather*}
By def\/inition, every~$\tau$-function can be determined by a~rational function in 7 initial variab\-les~$\tau_i$
$(i=1,2,\dots,7)$.
We note that the 7 initial variables are expressed by the~$\tau$-functions as the following (see
Fig.~\ref{fig:configuration_tau}):
\begin{gather}
\tau_1=\tau_1^{1,0},
\qquad
\tau_2=\tau_0^{1,1},
\qquad
\tau_3=\tau_1^{1,1},
\qquad
\tau_4=\tau_0^{0,0},
\nonumber
\\
\tau_5=\tau_1^{3,1},
\qquad
\tau_6=\tau_1^{2,1},
\qquad
\tau_7=\tau_0^{1,0}.
\label{eqn:taunmN_tau}
\end{gather}

\subsection[Hypergeometric~$\tau$-functions for the~$q$-P$_{\rm IV}$]{Hypergeometric
$\boldsymbol{\tau}$-functions for the~$\boldsymbol{q}$-P$\boldsymbol{_{\rm IV}}$}

The aim of this section is to construct the hypergeometric~$\tau$-functions for the~$q$-P$_{\rm IV}$.
We def\/ine the hypergeometric~$\tau$-functions for the~$q$-P$_{\rm IV}$ by $\tau_N^{n,m}$ consistent with the action of
$\langle R_{23},T_0\rangle$.
Here, we mean $\tau=\tau(\alpha)$ consistent with an action of transformation~$r$ as
\begin{gather*}
r.\tau=\tau(r.\alpha).
\end{gather*}

Hereinafter, we impose the condition~\eqref{eqn:condition_R23}, and then regard~$\tau$-functions $\tau_N^{n,m}$ as the
functions in $a_0$ and $a_2$ consistent with the action of $\langle R_{23},T_0\rangle$, i.e.,
\begin{gather*}
\tau_N^{n,m}=\tau_N^{0,0}\big(q^m a_0,q^{n/2}a_2\big).
\end{gather*}
By def\/inition, every~$\tau$-function $\tau_N^{n,m}$ is determined by a~rational function in $\tau_0^{n,m}$ and
$\tau_1^{n,m}$ (or $\tau_1,\dots,\tau_7$).
Thus, our purpose is determining $\tau_0^{n,m}$ and $\tau_1^{n,m}$ consistent with the action of $\langle
R_{23},T_0\rangle$ and constructing the closed-form expressions of $\tau_N^{n,m}$ $(N\geq2)$ under the condition
\begin{gather}
\label{eqn:qp4_condition}
a_0a_1=q,
\end{gather}
and the boundary condition
\begin{gather}
\label{eqn:qp4_boundary_condition}
\tau_N^{n,m}=0,
\qquad
N<0.
\end{gather}
Henceforth, we construct the hypergeometric~$\tau$-functions for the~$q$-P$_{\rm IV}$ in the following four steps.

\subsubsection*{Step 1.~Conditions of $\boldsymbol{\tau_0^{n,m}}$}

In the f\/irst step, we obtain the condition of $\tau_0^{n,m}$, which follows from the boundary
condition~\eqref{eqn:qp4_boundary_condition}.
\begin{lemma}
The following bilinear equations hold:
\begin{gather}
 \tau_{N+1}^{n,m}\tau_{N-1}^{n-1,m-1} -q^{(-n-4m+4N+7)/2}a_0^{-2}a_2^{-1}\tau_N^{n,m-1}\tau_N^{n-1,m}
\nonumber
\\
\qquad{}
-q^{(-n-2m+4N+4)/2}a_0^{-1}a_2^{-1}\tau_N^{n,m}\tau_N^{n-1,m-1}=0,
\label{eqn:bilinear_1}
\\
 \tau_{N+1}^{n,m}\tau_{N-1}^{n-1,m} +q^{2N-n+1}a_2^{-2}\big(q^{(-n+2N+1)/2}a_2^{-1}-1\big)\tau_N^{n-1,m}\tau_N^{n,m}
\nonumber
\\
\qquad{}
-q^{(-3n+6N+3)/2}a_2^{-3}\tau_N^{n-2,m}\tau_N^{n+1,m}=0,
\label{eqn:bilinear_2}
\\
 \tau_{N+1}^{n,m}\tau_{N-1}^{n,m} +q^{(-2n+6N+1)/2}a_2^{-2}\big(1-q^{N-m+1}a_0^{-1}\big)\big(\tau_N^{n,m}\big)^2
\nonumber
\\
\qquad{}
-q^{4N-4m+4}a_0^{-4}\tau_N^{n,m-1}\tau_N^{n,m+1}=0.
\label{eqn:bilinear_3}
\end{gather}
\end{lemma}
\begin{proof}
Application of $T_1$ on $\tau_3$ yields the following bilinear equations:
\begin{gather}
 T_1(\tau_3)\tau_4 -q^2a_0^{-1}a_1a_2^{-1}\tau_1R_{23}^{-1}(\tau_3)-q^{3/2}a_1a_2^{-1}\tau_3R_{23}^{-1}(\tau_1)=0,
\label{eqn:bilinear_1_proof}
\\
 T_1(\tau_3)\tau_2 +q^{3/2}a_0a_1a_2^{-2}(1-a_1)(\tau_3)^2 -a_1^4\tau_1T_0(\tau_3)=0.
\label{eqn:bilinear_3_proof}
\end{gather}
Applying $R_{23}^{n-1}T_0^{m-1}T_1^{N-1}$ on equations~\eqref{eqn:bilinear_1_proof}
and~\eqref{eqn:bilinear_3_proof} and substituting condition~\eqref{eqn:qp4_condition} in them, we obtain
equations~\eqref{eqn:bilinear_1} and~\eqref{eqn:bilinear_3}, respectively.
Similarly, application of $T_1$ on $\tau_6$ yields
\begin{gather}
\label{eqn:bilinear_2_proof}
T_1(\tau_6)\tau_2 +qa_2^{-2}\big(q^{1/2}a_2^{-1}-1\big)\tau_3\tau_6 -q^{3/2}a_2^{-3}R_{23}^{-1}(\tau_3)\tau_5=0.
\end{gather}
Then, applying $R_{23}^{n-2}T_0^{m-1}T_1^{N-1}$ on equation~\eqref{eqn:bilinear_2_proof} and substituting
condition~\eqref{eqn:qp4_condition} in it, we obtain equation~\eqref{eqn:bilinear_2}.
Although we do not write the action of $R_{23}$ and $T_0$ on the variables $\tau_i$ here, it will be described in the
next step.
\end{proof}

Putting $N=0$ in equations~\eqref{eqn:bilinear_1}--\eqref{eqn:bilinear_3} and using the boundary
condition~\eqref{eqn:qp4_boundary_condition}, we get the following conditions:
\begin{gather}
 \frac{\tau_{0}^{n+1,m}\tau_{0}^{n,m+1}}{\tau_{0}^{n,m}\tau_{0}^{n+1,m+1}} =-q^{(2m-1)/2}a_0,
\label{eqn:bilinear_1_2}
\\
 \frac{\tau_{0}^{n,m}\tau_{0}^{n+3,m}}{\tau_{0}^{n+1,m}\tau_{0}^{n+2,m}} =1-q^{(n+1)/2}a_2,
\\
 \frac{\tau_{0}^{n,m}\tau_0^{n,m+2}}{\big(\tau_{0}^{n,m+1}\big)^2} =q^{(-2n+8m+1)/2}a_0^4a_2^{-2}\big(1-q^{-m}a_0^{-1}\big).
\label{eqn:bilinear_3_2}
\end{gather}

\subsubsection*{Step 2.\ Conditions of $\boldsymbol{\tau_1^{n,m}}$}

In the second step, we shall get the conditions of $\tau_1^{n,m}$ from the consistency with the action of $\langle
R_{23},T_0\rangle$.
By def\/initions~\eqref{eqn:translations} and~\eqref{eqn:R23} and Proposition~\ref{prop:weyl_group_tau}, the action of
$T_0$ and $R_{23}$ are given by the follows:
\begin{gather}
 T_0(\tau_1)=\tau_3,
\qquad
T_0(\tau_7)=\tau_2,
\qquad
R_{23}(\tau_3)=\tau_6,
\qquad
R_{23}(\tau_4)=\tau_7,
\qquad
R_{23}(\tau_6)=\tau_5,
\nonumber
\\
 T_0(\tau_2) =\frac{q^{3/2}a_0a_1^{-1}a_2^{-1}\tau_2T_0(\tau_6)+qa_0^2a_2^{-1}T_0(\tau_4)T_0(\tau_5)} {\tau_6},
\label{eqn:T0_tau2}
\\
 T_0(\tau_3) =\frac{a_0^2a_1T_0(\tau_4)T_0(\tau_5)+q^{-1/2}a_0a_1a_2T_0(\tau_6)\tau_2} {R_{23}(\tau_2)},
\\
 T_0(\tau_4) =\frac{a_2\tau_2\tau_6+qa_0^{-1}a_1^{-2}\tau_3R_{23}(\tau_2)}{\tau_5},
\\
 T_0(\tau_5)
=\frac{qa_0^{-2}a_1^{-2}a_2^{-1}\tau_2\tau_6+q^{3/2}a_0^{-2}a_1^{-3}a_2^{-1}\tau_3R_{23}(\tau_2)} {\tau_4},
\\
 T_0(\tau_6) =\frac{qa_1^{-2}a_2^{-1}\tau_2\tau_6+q^{3/2}a_0^{-1}a_1^{-3}a_2^{-1}\tau_3R_{23}(\tau_2)}
{\tau_7},
\\
 T_0^{-1}(\tau_1) =\frac{q^{-2}a_0^2a_1a_2^{-1}T_0^{-1}(\tau_4)T_0^{-1}(\tau_5)
+q^{-1/2}a_0a_1a_2^{-1}T_0^{-1}(\tau_6)T_0^{-1}(\tau_7)} {R_{23}(\tau_7)},
\\
 T_0^{-1}(\tau_4) =\frac{qa_0^{-2}a_1^{-2}a_2^{-1}\tau_7T_0^{-1}(\tau_6)
+q^{1/2}a_0^{-2}a_1^{-3}a_2^{-1}\tau_1R_{23}(\tau_7)} {\tau_5},
\\
 T_0^{-1}(\tau_5) =\frac{a_2\tau_7T_0^{-1}(\tau_6)+a_0^{-1}a_1^{-2}\tau_1R_{23}(\tau_7)} {\tau_4},
\\
 T_0^{-1}(\tau_6) =\frac{q^{-1}a_0^2a_2^{-1}\tau_4\tau_5+q^{-1/2}a_0a_1^{-1}a_2^{-1}\tau_6\tau_7} {\tau_2},
\\
 T_0^{-1}(\tau_7) =\frac{q^{-1}a_1^{-2}a_2^{-1}\tau_7T_0^{-1}(\tau_6)
+q^{-1/2}a_0^{-1}a_1^{-3}a_2^{-1}\tau_1R_{23}(\tau_7)} {\tau_6},
\\
 R_{23}(\tau_1) =\frac{q^{-1}a_0^2a_2^{-1}\tau_4\tau_5+q^{-1/2}a_0a_1^{-1}a_2^{-1}\tau_6\tau_7} {\tau_2},
\\
 R_{23}(\tau_2)=\frac{q^{-1}a_0^2a_1a_2^{-1}\tau_4\tau_5+q^{-1/2}a_0a_1a_2^{-1}\tau_6\tau_7} {\tau_1},
\\
 R_{23}(\tau_5) =\frac{q^{3/2}a_0^{-2}a_2^2R_{23}(\tau_1)R_{23}(\tau_2)+qa_0^{-2}a_1^{-1}\tau_6R_{23}(\tau_7)}
{\tau_4},
\\
 R_{23}(\tau_7) =\frac{q^{-1}a_0^2a_1\tau_4\tau_5+q^{-1/2}a_0a_1a_2\tau_6\tau_7}{\tau_3},
\\
 R_{23}^{-1}(\tau_1) =\frac{q^{-1/2}a_0^2a_1a_2^{-1}R_{23}^{-1}(\tau_4)\tau_6+a_0a_1a_2^{-1}\tau_3\tau_4}
{\tau_2},
\\
 R_{23}^{-1}(\tau_2) =\frac{q^{-1/2}a_0^2a_2^{-1}R_{23}^{-1}(\tau_4)\tau_6+a_0a_1^{-1}a_2^{-1}\tau_3\tau_4}
{\tau_1},
\\
 R_{23}^{-1}(\tau_3) =\frac{q^{-1}a_0^2a_1R_{23}^{-1}(\tau_4)\tau_6+q^{-1}a_0a_1a_2\tau_3\tau_4} {\tau_7},
\\
 R_{23}^{-1}(\tau_4) =\frac{q^{1/2}a_0^{-2}a_2^2\tau_1\tau_2+qa_0^{-2}a_1^{-1}\tau_3\tau_7}{\tau_5}.
\label{eqn:R23^-1_tau4}
\end{gather}
Using notation~\eqref{eqn:taunmN_tau} and condition~\eqref{eqn:qp4_condition}, we can rewrite
equations~\eqref{eqn:T0_tau2}--\eqref{eqn:R23^-1_tau4}~as
\begin{gather}
 a_2\tau^{2,1}_1\tau^{1,2}_0 =q^{1/2}a_0^2\tau^{1,1}_0\tau^{2,2}_1+qa_0^2\tau^{0,1}_0\tau^{3,2}_1,
\label{eqn:T0_tau2_2}
\\
 \tau^{2,1}_0\tau^{1,2}_1 =qa_0\tau^{0,1}_0\tau^{3,2}_1+q^{1/2}a_2\tau^{2,2}_1\tau^{1,1}_0,
\label{eqn:T0_tau3_2}
\\
 q\tau^{3,1}_1\tau^{0,1}_0 =qa_2\tau^{1,1}_0\tau^{2,1}_1+a_0\tau^{1,1}_1\tau^{2,1}_0,
\label{eqn:T0_tau4_2}
\\
 q^{3/2}a_2\tau^{0,0}_0\tau^{3,2}_1 =q^{1/2}\tau^{1,1}_0\tau^{2,1}_1+a_0\tau^{1,1}_1\tau^{2,1}_0,
\label{eqn:T0_tau5_2}
\\
 q^{3/2}a_2\tau^{1,0}_0\tau^{2,2}_1 =q^{1/2}a_0^2\tau^{1,1}_0\tau^{2,1}_1+a_0^2\tau^{1,1}_1\tau^{2,1}_0,
\label{eqn:T0_tau6_2}
\\
 a_2\tau^{2,0}_0\tau^{1,-1}_1 =q^{-1}a_0\tau^{0,-1}_0\tau^{3,0}_1+q^{1/2}\tau^{2,0}_1\tau^{1,-1}_0,
\label{eqn:T0^-1_tau1_2}
\\
 q^{5/2}a_2\tau^{3,1}_1\tau^{0,-1}_0 =q^{3/2}\tau^{1,0}_0\tau^{2,0}_1+a_0\tau^{1,0}_1\tau^{2,0}_0,
\label{eqn:T0^-1_tau4_2}
\\
 q^2\tau^{0,0}_0\tau^{3,0}_1 =q^2a_2\tau^{1,0}_0\tau^{2,0}_1+a_0\tau^{1,0}_1\tau^{2,0}_0,
\label{eqn:T0^-1_tau5_2}
\\
 q^{3/2}a_2\tau^{1,1}_0\tau^{2,0}_1 =q^{1/2}a_0^2\tau^{0,0}_0\tau^{3,1}_1+a_0^2\tau^{2,1}_1\tau^{1,0}_0,
\label{eqn:T0^-1_tau6_2}
\\
 q^{7/2}a_2\tau^{2,1}_1\tau^{1,-1}_0 =q^{1/2}a_0^2\tau^{1,0}_0\tau^{2,0}_1+a_0^2\tau^{1,0}_1\tau^{2,0}_0,
\label{eqn:T0^-1_tau7_2}
\\
 q^{3/2}a_2\tau^{1,1}_0\tau^{2,0}_1 =q^{1/2}a_0^2\tau^{0,0}_0\tau^{3,1}_1+a_0^2\tau^{2,1}_1\tau^{1,0}_0,
\label{eqn:R23_tau1_2}
\\
 a_2\tau^{1,0}_1\tau^{2,1}_0 =a_0\tau^{0,0}_0\tau^{3,1}_1+q^{1/2}\tau^{2,1}_1\tau^{1,0}_0,
\label{eqn:R23_tau2_2}
\\
 a_0^2\tau^{0,0}_0\tau^{4,1}_1 =q^{3/2}a_2^2\tau^{2,0}_1\tau^{2,1}_0+a_0\tau^{2,1}_1\tau^{2,0}_0,
\label{eqn:R23_tau5_2}
\\
 \tau^{1,1}_1\tau^{2,0}_0 =a_0\tau^{0,0}_0\tau^{3,1}_1+q^{1/2}a_2\tau^{2,1}_1\tau^{1,0}_0,
\label{eqn:R23_tau7_2}
\\
 a_2\tau^{1,1}_0\tau^{0,0}_1 =q^{1/2}a_0\tau^{-1,0}_0\tau^{2,1}_1+q\tau^{1,1}_1\tau^{0,0}_0,
\label{eqn:R23^-1_tau1_2}
\\
 qa_2\tau^{1,0}_1\tau^{0,1}_0 =q^{1/2}a_0^2\tau^{-1,0}_0\tau^{2,1}_1+a_0^2\tau^{1,1}_1\tau^{0,0}_0,
\label{eqn:R23^-1_tau2_2}
\\
 \tau^{1,0}_0\tau^{0,1}_1 =a_0\tau^{-1,0}_0\tau^{2,1}_1+a_2\tau^{1,1}_1\tau^{0,0}_0,
\label{eqn:R23^-1_tau3_2}
\\
 a_0^2\tau^{3,1}_1\tau^{-1,0}_0 =q^{1/2}a_2^2\tau^{1,0}_1\tau^{1,1}_0+a_0\tau^{1,1}_1\tau^{1,0}_0,
\label{eqn:R23^-1_tau4_2}
\end{gather}
respectively.
By setting
\begin{gather}
\label{eqn:gage_qp4_tau1}
\tau^{n,m}_1=\big(q^{(n-1)/2}a_2;q^{1/2}\big)_\infty\tau^{n,m}_0F_{n,m},
\end{gather}
and using conditions~\eqref{eqn:bilinear_1_2}--\eqref{eqn:bilinear_3_2},
equations~\eqref{eqn:T0_tau2_2}--\eqref{eqn:R23^-1_tau4_2} can be reduced to the following contiguity relations:
\begin{gather}
 F_{n+2,m}-q^{(n-1)/2}a_2F_{n+1,m}-q^{m-2}a_0\big(1-q^{(n-1)/2}a_2\big)F_{n,m}=0,
\label{eqn:relation_F_1}
\\
 F_{n+1,m+1}-q^{m-1}a_0F_{n,m+1}-q^{(n-2)/2}a_2F_{n,m}=0,
\label{eqn:relation_F_2}
\\
 q^{1/2}F_{n+2,m+1}-q^{1/2}F_{n+1,m+1}+q^{(n-1)/2}a_2\big(1-q^{(n-1)/2}a_2\big)F_{n,m}=0,
\label{eqn:relation_F_3}
\\
 \big(1-q^{m-1}a_0\big)F_{n+1,m+1}-q^{(n-1)/2}a_2F_{n+1,m} -q^{n/2-1}a_2\big(1-q^{(n-1)/2}a_2\big)F_{n,m}=0,
\label{eqn:relation_F_4}
\\
 q^{3/2}\big(1-q^{m-1}a_0\big)F_{n+2,m+1}-q^{(n+2)/2}a_2F_{n+1,m} \nonumber\\
 \qquad{} -q^{(2m+n-1)/2}a_0a_2\big(1-q^{(n-1)/2}a_2\big)F_{n,m}=0,
\label{eqn:relation_F_5}
\\
 qF_{n+2,m+1}-q^ma_0F_{n,m+1}-q^{(2n-1)/2}a_2^2F_{n,m}=0.
\label{eqn:relation_F_6}
\end{gather}
The correspondence between equations~\eqref{eqn:T0_tau2_2}--\eqref{eqn:R23^-1_tau4_2} and
equations~\eqref{eqn:relation_F_1}--\eqref{eqn:relation_F_6} is established as follows:
\begin{gather*}
 \eqref{eqn:T0_tau3_2},~\eqref{eqn:T0_tau4_2},~\eqref{eqn:T0^-1_tau5_2},~\eqref{eqn:R23_tau7_2},~\eqref{eqn:R23^-1_tau3_2}
\Rightarrow \eqref{eqn:relation_F_1},
\\
 \eqref{eqn:T0_tau2_2},~\eqref{eqn:T0^-1_tau6_2},~\eqref{eqn:R23_tau1_2},~\eqref{eqn:R23^-1_tau2_2}
\Rightarrow \eqref{eqn:relation_F_2},
\\
 \eqref{eqn:T0^-1_tau1_2},~\eqref{eqn:R23_tau2_2},~\eqref{eqn:R23^-1_tau1_2} \Rightarrow \eqref{eqn:relation_F_3},
\\
 \eqref{eqn:T0_tau6_2},~\eqref{eqn:T0^-1_tau7_2} \Rightarrow \eqref{eqn:relation_F_4},
\\
 \eqref{eqn:T0_tau5_2},~\eqref{eqn:T0^-1_tau4_2} \Rightarrow \eqref{eqn:relation_F_5},
\\
 \eqref{eqn:R23_tau5_2},~\eqref{eqn:R23^-1_tau4_2} \Rightarrow \eqref{eqn:relation_F_6}.
\end{gather*}

\subsubsection*{Step 3.\ Determination of $\boldsymbol{\tau_0^{n,m}}$ and
$\boldsymbol{\tau_1^{n,m}}$}

In this step, we determine $\tau_0^{n,m}$ and $\tau_1^{n,m}$, i.e., we solve
equations~\eqref{eqn:bilinear_1_2}--\eqref{eqn:bilinear_3_2} and
equations~\eqref{eqn:relation_F_1}--\eqref{eqn:relation_F_6}.
It is easily verif\/ied that the function
\begin{gather}
\tau_0^{n,m} = \big(q^{m}a_0;q,q\big)_{\infty}\big(q^{(n+1)/2}a_2;q^{1/2},q\big)_{\infty}
\Gamma\big(q^{(2n+2m-3)/4}a_0^{1/2}a_2;q^{1/2},q^{1/2}\big)
\nonumber
\\
\phantom{\tau_0^{n,m} =}{}
 \times\frac{\Gamma\big(q^{(n-m+1)/4}a_0^{-1/4}a_2^{1/2};q^{1/4},q^{1/4}\big)
\Gamma\big(q^{(n-m)/4}a_0^{-1/4}a_2^{1/2};q^{1/4},q^{1/4}\big)}
{\Gamma\big({-}q^{3m-1}a_0^{3};q^3,q^3\big)\Gamma\big({-}q^{2n}a_2^4;q^2,q^2\big)},
\label{eqn:gage_qp4_tau0}
\end{gather}
is the solution of equations~\eqref{eqn:bilinear_1_2}--\eqref{eqn:bilinear_3_2}.
Therefore, the aim of this step is to solve the equations~\eqref{eqn:relation_F_1}--\eqref{eqn:relation_F_6}.
Since equations~\eqref{eqn:relation_F_1}--\eqref{eqn:relation_F_6} are overdetermined system, let us f\/irst consider the
essential conditions of $F_{n,m}$.
\begin{lemma}
\label{lemma:Fnm_only2}
Equations~\eqref{eqn:relation_F_1} and~\eqref{eqn:relation_F_2} are essential conditions for $F_{n,m}$.
\end{lemma}
\begin{proof}
Eliminating $F_{n,m+1}$ from equations~\eqref{eqn:relation_F_1}$_{m\to m+1}$ and~\eqref{eqn:relation_F_2}, we obtain
equation~\eqref{eqn:relation_F_3}.
In a~similar manner, equations~\eqref{eqn:relation_F_4}--\eqref{eqn:relation_F_6} can be proven by the following
procedures: elimi\-na\-ting $F_{n+2,m+1}$ from equations~\eqref{eqn:relation_F_2}$_{n\to n+1}$ and~\eqref{eqn:relation_F_3},
we obtain equation~\eqref{eqn:relation_F_4}; elimi\-na\-ting $F_{n+1,m+1}$ from equations~\eqref{eqn:relation_F_2}$_{n\to
n+1}$ and~\eqref{eqn:relation_F_3}, we obtain equation~\eqref{eqn:relation_F_5}; eliminating $F_{n+1,m+1}$ from
equations~\eqref{eqn:relation_F_2} and~\eqref{eqn:relation_F_3}, we obtain equation~\eqref{eqn:relation_F_6}.
These calculations mean that if $F_{n,m}$ satisf\/ies conditions~\eqref{eqn:relation_F_1} and~\eqref{eqn:relation_F_2},
then it also satisf\/ies conditions~\eqref{eqn:relation_F_3}--\eqref{eqn:relation_F_6}.
Therefore we have completed the proof.
\end{proof}

Next, we solve equations~\eqref{eqn:relation_F_1} and~\eqref{eqn:relation_F_2}.
\begin{lemma}
\label{lemma:sol_F1_F2}
The general solution of contiguity relations~\eqref{eqn:relation_F_1} and~\eqref{eqn:relation_F_2} is given~by
\begin{gather*}
F_{n,m}= A_{n,m}\frac{\Theta\big(q^{n/2}a_2;q^{1/2}\big) \Theta\big(q^{(2m-1)/2}a_0;q\big)\big(q^{(m-1)/2}a_0^{1/2};q^{1/2}\big)_\infty}
{\Theta\big(q^{(n+m-2)/2}a_0^{1/2}a_2;q^{1/2}\big)}
\nonumber
\\
\phantom{F_{n,m}=+}{}
\times{}_2\varphi_1\left(
\begin{matrix}
0,q^{(-m+2)/2}a_0^{-1/2}
\\
-q^{1/2}
\end{matrix}
;q^{1/2},q^{(n-1)/2}a_2\right)
\nonumber
\\
\phantom{F_{n,m}=}{}
 +B_{n,m}\frac{\Theta\big(q^{n/2}a_2;q^{1/2}\big)\Theta\big(q^{(2m-1)/2}a_0;q\big) \big({-}q^{(m-1)/2}a_0^{1/2};q^{1/2}\big)_\infty}
{\Theta\big({-}q^{(n+m-2)/2}a_0^{1/2}a_2;q^{1/2}\big)}
\nonumber
\\
\phantom{F_{n,m}=+}
\times{}_2\varphi_1\left(
\begin{matrix}
0,-q^{(-m+2)/2}a_0^{-1/2}
\\
-q^{1/2}
\end{matrix}
;q^{1/2},q^{(n-1)/2}a_2\right),
\end{gather*}
where $A_{n,m}$ and $B_{n,m}$ are periodic functions of period one for~$n$ and~$m$, i.e.,
\begin{gather*}
A_{n,m}=A_{n+1,m}=A_{n,m+1},
\qquad
B_{n,m}=B_{n+1,m}=B_{n,m+1}.
\end{gather*}
\end{lemma}

\begin{proof}
For convenience, we put
\begin{gather*}
t=q^{n/2}a_2,
\qquad
\alpha=q^ma_0,
\qquad
F_{n,m}=F(t,\alpha).
\end{gather*}
Then, equations~\eqref{eqn:relation_F_1} and~\eqref{eqn:relation_F_2} can be rewritten as
\begin{gather}
 F(qt,\alpha)-q^{-1/2}tF\big(q^{1/2}t,\alpha\big) -q^{-2}\alpha\big(1-q^{-1/2}t\big)F(t,\alpha)=0,
\label{eqn:relation_F_1_2}
\\
 F\big(q^{1/2}t,q\alpha\big)-q^{-1}\alpha F(t,q\alpha) -q^{-1}tF(t,\alpha)=0,
\label{eqn:relation_F_2_2}
\end{gather}
respectively.
Substituting
\begin{gather*}
F(t,\alpha)=D(t,\alpha)\sum\limits^{\infty}_{k=0}C_k(\alpha)t^k,
\end{gather*}
in equation~\eqref{eqn:relation_F_1_2}, we obtain
\begin{gather}
 D(qt,\alpha)=q^{-2}\alpha D(t,\alpha),
\label{eqn:D_first_order}
\\
 C_k(\alpha)=\frac{\big(q^2D\big(q^{1/2}t,\alpha\big)D(t,\alpha)^{-1}\alpha^{-1};q^{1/2}\big)_k}
{q^{k/2}\big({-}q^{1/2},q^{1/2};q^{1/2}\big)_k} C_0(\alpha).
\nonumber
\end{gather}
Therefore, we obtain the solution of equation~\eqref{eqn:relation_F_1_2}:
\begin{gather}
F(t,\alpha) =D_1(t,\alpha)\, {}_2\varphi_1\left(
\begin{matrix}
0,q\alpha^{-1/2}
\\
-q^{1/2}
\end{matrix}
;q^{1/2},q^{-1/2}t\right)
\nonumber
\\
\phantom{F(t,\alpha) =}{}
+D_2(t,\alpha)\, {}_2\varphi_1\left(
\begin{matrix}
0,-q\alpha^{-1/2}
\\
-q^{1/2}
\end{matrix}
;q^{1/2},q^{-1/2}t\right),
\label{eqn:proof_F_Hyper_1}
\end{gather}
where $D_1(t,\alpha)$ and $D_2(t,\alpha)$ are the solutions of~\eqref{eqn:D_first_order} which satisfy
\begin{gather}
 D_1\big(q^{1/2}t,\alpha\big)=q^{-1}\alpha^{1/2}D_1(t,\alpha),
\label{eqn:proof_F_Hyper_2}
\\
 D_2\big(q^{1/2}t,\alpha\big)=-q^{-1}\alpha^{1/2}D_2(t,\alpha),
\label{eqn:proof_F_Hyper_3}
\end{gather}
respectively.
Substituting~\eqref{eqn:proof_F_Hyper_1} in equation~\eqref{eqn:relation_F_2_2}, we can obtain the following equations:
\begin{gather}
 q^{-1/2}\alpha^{1/2} {}_2\varphi_1\left(
\begin{matrix}
0,q^{1/2}\alpha^{-1/2}
\\
-q^{1/2}
\end{matrix}
;q^{1/2},t\right) -q^{-1}\alpha\, {}_2\varphi_1\left(
\begin{matrix}
0,q^{1/2}\alpha^{-1/2}
\\
-q^{1/2}
\end{matrix}
;q^{1/2},q^{-1/2}t\right)
\nonumber
\\
\qquad{}
-q^{-1}t\frac{D_1(t,\alpha)}{D_1(t,q\alpha)}\, {}_2\varphi_1\left(
\begin{matrix}
0,q\alpha^{-1/2}
\\
-q^{1/2}
\end{matrix}
;q^{1/2},q^{-1/2}t\right)=0,
\label{eqn:proof_F_Hyper_4}
\\
 q^{-1/2}\alpha^{1/2} {}_2\varphi_1\left(
\begin{matrix}
0,-q^{1/2}\alpha^{-1/2}
\\
-q^{1/2}
\end{matrix}
;q^{1/2},t\right) +q^{-1}\alpha\, {}_2\varphi_1\left(
\begin{matrix}
0,-q^{1/2}\alpha^{-1/2}
\\
-q^{1/2}
\end{matrix}
;q^{1/2},q^{-1/2}t\right)
\nonumber
\\
\qquad{}
+q^{-1}t\frac{D_2(t,\alpha)}{D_2(t,q\alpha)}\, {}_2\varphi_1\left(
\begin{matrix}
0,-q\alpha^{-1/2}
\\
-q^{1/2}
\end{matrix}
;q^{1/2},q^{-1/2}t\right)=0.
\label{eqn:proof_F_Hyper_5}
\end{gather}
By the def\/inition of basic hypergeometric series ${}_2\varphi_1$, it is easily verif\/ied that
\begin{gather}
{}_2\varphi_1\left(
\begin{matrix}
0,a
\\
c
\end{matrix}
;q^{1/2},z\right) -a^{-1}\,{}_2\varphi_1\left(
\begin{matrix}
0,a
\\
c
\end{matrix}
;q^{1/2},q^{-1/2}z\right)
\nonumber
\\
\qquad{}
-\big(1-a^{-1}\big)\,{}_2\varphi_1\left(
\begin{matrix}
0,q^{1/2}a
\\
c
\end{matrix}
;q^{1/2},q^{-1/2}z\right)=0.
\label{eqn:transform_2phi1_1}
\end{gather}
Therefore, we obtain the following conditions from equations~\eqref{eqn:proof_F_Hyper_4} and~\eqref{eqn:proof_F_Hyper_5}
by using equation~\eqref{eqn:transform_2phi1_1}:
\begin{gather}
 D_1(t,q\alpha) =-\frac{t}{\alpha(1-q^{1/2}\alpha^{-1/2})} D_1(t,\alpha),
\label{eqn:proof_F_Hyper_6}
\\
 D_2(t,q\alpha) =-\frac{t}{\alpha(1+q^{1/2}\alpha^{-1/2})} D_2(t,\alpha).
\label{eqn:proof_F_Hyper_7}
\end{gather}
By setting
\begin{gather*}
 D_1(t,\alpha)=\frac{\Theta\big(t;q^{1/2}\big)\Theta\big(q^{-1/2}\alpha;q\big)\big(q^{-1/2}\alpha^{1/2};q^{1/2}\big)_\infty}
{\Theta\big(q^{-1}\alpha^{1/2}t;q^{1/2}\big)} A(t,\alpha),
\\
 D_2(t,\alpha)=\frac{\Theta\big(t;q^{1/2}\big)\Theta\big(q^{-1/2}\alpha;q\big)\big({-}q^{-1/2}\alpha^{1/2};q^{1/2}\big)_\infty}
{\Theta\big({-}q^{-1}\alpha^{1/2}t;q^{1/2}\big)} B(t,\alpha),
\end{gather*}
equations~\eqref{eqn:proof_F_Hyper_2},~\eqref{eqn:proof_F_Hyper_3},~\eqref{eqn:proof_F_Hyper_6}
and~\eqref{eqn:proof_F_Hyper_7} can be rewritten as
\begin{alignat*}{3}
& A\big(q^{1/2}t,\alpha\big)=A(t,\alpha),
\qquad &&
 B\big(q^{1/2}t,\alpha\big)=B(t,\alpha),&
\\
& A(t,q\alpha)=A(t,\alpha),
\qquad&&
 B(t,q\alpha)=B(t,\alpha),&
\end{alignat*}
respectively.
This completes the proof.
\end{proof}

It was shown that hypergeometric solutions of a~symmetric discrete Painlev\'e equation, which can be obtained~by
projective reduction, have two expressions and there are the following dif\/ferences between
the two expressions (see~\cite[Section 2.3]{KNT2011:MR2773334}):
\begin{enumerate}\itemsep=0pt
\item[(i)] the bases of hypergeometric series appearing in the solutions are dif\/ferent;
\item[(ii)] the periodicities of periodic functions appearing in the solutions are dif\/ferent.
\end{enumerate}
The dif\/ferences between these two expressions can be explained by the factorization of the linear dif\/ference operators
associated with the three-term relation of the hypergeometric functions (see~\cite[Section~3.2]{KNT2011:MR2773334}).
Namely, we can see these dif\/ferences by comparing Lemmas~\ref{lemma:sol_F1_F2} and~\ref{lemma:sol_G2_G123}.
To get another expression, we f\/irst reselect essential conditions of $F_{n,m}$.
\begin{lemma}
Equations~\eqref{eqn:relation_F_2} and
\begin{gather}
 q^{m-1}a_0\big(1-q^ma_0\big)F_{n,m+2} -q^{(n-5)/2}a_2\big(\big(1+q^{1/2}\big)q^ma_0-q^{n/2}a_2\big)F_{n,m+1}
\nonumber
\\
\qquad{}
 -q^{(2n-5)/2}a_2^2F_{n,m}=0,
\label{eqn:relation_F_123}
\end{gather}
are essential conditions of $F_{n,m}$.
\end{lemma}
\begin{proof}
Eliminating $F_{n,m}$ from equations~\eqref{eqn:relation_F_2} and~\eqref{eqn:relation_F_123}, we obtain
\begin{gather}
q^{m-2}a_0\big(1-q^{m-1}a_0\big)F_{n,m+1}-q^{(n-3)/2}a_2F_{n+1,m}
\nonumber
\\
\qquad{}
-q^{(n-5)/2}a_2\big(q^{(2m-1)/2}a_0-q^{n/2}a_2\big)F_{n,m}=0.
\label{eqn:relation_F_1232}
\end{gather}
Similarly, eliminating $F_{n,m+1}$ from equations~\eqref{eqn:relation_F_2} and~\eqref{eqn:relation_F_1232}, we obtain
\begin{gather}
 \big(1-q^{m-1}a_0\big)F_{n+1,m+1}-q^{(n-1)/2}a_2F_{n+1,m} -q^{(n-2)/2}a_2\big(1-q^{(n-1)/2}a_2\big)F_{n,m}=0.
\label{eqn:relation_F_12322}
\end{gather}
Finally, eliminating $F_{n+1,m+1}$ from equations~\eqref{eqn:relation_F_1232}$_{n\to n+1}$
and~\eqref{eqn:relation_F_12322}, we obtain equation~\eqref{eqn:relation_F_1}.
This result together with Lemma~\ref{lemma:Fnm_only2} complete the proof.
\end{proof}

By setting
\begin{gather}
\label{define:Gnm}
F_{n,m} =\frac{\Theta\big(q^ma_0;q\big)\Theta\big(q^{n/2}a_2;q^{1/2}\big)} {\Theta\big({-}q^{n/2}a_2;q^{1/2}\big)} G_{n-3,m-1},
\end{gather}
equations~\eqref{eqn:relation_F_2} and~\eqref{eqn:relation_F_123} can be rewritten as
\begin{gather}
 q^{-m+1}a_0^{-1}G_{n-2,m}+G_{n-3,m}-q^{n/2}a_2G_{n-3,m-1}=0,
\label{eqn:relation_G_2}
\\
 G_{n,m-2}+\big(q^{-m+1}a_0^{-1}-\big(1+q^{1/2}\big)q^{(-n-3)/2}a_2^{-1}\big)G_{n,m-1}
\nonumber
\\
\qquad{}
-q^{(-2n-5)/2}a_2^{-2}\big(q^{-m+1}a_0^{-1}-1\big)G_{n,m}=0,
\label{eqn:relation_G_123}
\end{gather}
respectively.
Before solving equations~\eqref{eqn:relation_G_2} and~\eqref{eqn:relation_G_123}, we prepare the following lemma:
\begin{lemma}
\label{lemma:3term_2phi1}
The following recurrence relations hold:
\begin{gather}
 {}_2\varphi_1\left(
\begin{matrix}
a,b
\\
c
\end{matrix}
;q,z\right) -{}_2\varphi_1\left(
\begin{matrix}
a,b
\\
c
\end{matrix}
;q,qz\right) =\frac{(1-a)(1-b)z}{1-c}\,{}_2\varphi_1\left(
\begin{matrix}
qa,qb
\\
qc
\end{matrix}
;q,z\right),
\label{eqn:hyper_relation_1}
\\
 \big(q^{-1}c-1\big)\, {}_2\varphi_1\left(
\begin{matrix}
a,b
\\
q^{-1}c
\end{matrix}
;q,z\right) +{}_2\varphi_1\left(
\begin{matrix}
a,b
\\
c
\end{matrix}
;q,z\right) -q^{-1}c\, {}_2\varphi_1\left(
\begin{matrix}
a,b
\\
c
\end{matrix}
;q,qz\right)=0.
\label{eqn:hyper_relation_2}
\end{gather}
\end{lemma}
\begin{proof}
Substituting
\begin{gather*}
{}_2\varphi_1\left(
\begin{matrix}
a,b
\\
c
\end{matrix}
;q,z\right) =1+\sum\limits_{n=0}^{\infty}\frac{(qa,qb;q)_n}{(qc,q;q)_n} \frac{(1-a)(1-b)}{(1-c)\big(1-q^{n+1}\big)} z^{n+1},
\end{gather*}
in the left-hand side of equation~\eqref{eqn:hyper_relation_1}, we obtain the right-hand side.
Equation~\eqref{eqn:hyper_relation_2} can be easily verif\/ied as the following:
\begin{gather*}
{}_2\varphi_1\left(
\begin{matrix}
a,b
\\
q^{-1}c
\end{matrix}
;q,z\right) = \sum\limits_{n=0}^{\infty}\frac{(a,b;q)_n}{(c,q;q)_n} \frac{1-q^{n-1}c}{1-q^{-1}c} z^n
\nonumber
\\
\phantom{{}_2\varphi_1\left(\begin{matrix}a,b\\q^{-1}c\end{matrix};q,z\right)}
= \frac{1}{1-q^{-1}c}\, {}_2\varphi_1\left(
\begin{matrix}
a,b
\\
c
\end{matrix}
;q,z\right) -\frac{q^{-1}c}{1-q^{-1}c}\, {}_2\varphi_1\left(
\begin{matrix}
a,b
\\
c
\end{matrix}
;q,qz\right).
\end{gather*}
Therefore we have completed the proof.
\end{proof}

Using Lemma~\ref{lemma:3term_2phi1}, we obtain the following lemma:
\begin{lemma}
\label{lemma:sol_G2_G123}
The general solution of equations~\eqref{eqn:relation_G_2} and~\eqref{eqn:relation_G_123} is given~by
\begin{gather*}
G_{n,m} = \Lambda_{n,m}\frac{\Theta\big(q^{(n-2m+2)/2}a_0^{-1}a_2;q\big)}
{\Theta\big(q^{-m}a_0^{-1};q\big)\Theta\big(q^{n/2}a_2;q\big)\big(q^{(n+3)/2}a_2;q\big)_\infty\big(q^{-1/2};q\big)_\infty}
\\
\phantom{G_{n,m} =+}{}
\times {}_2\varphi_1\left(
\begin{matrix}
0,q^{(n+3)/2}a_2
\\
q^{3/2}
\end{matrix}
;q,q^{-m+1}a_0^{-1}\right)
\\
\phantom{G_{n,m} =}{}
 +\Lambda_{n+1,m}\frac{q^{1/2}\Theta\big(q^{(n-2m+3)/2}a_0^{-1}a_2;q\big)}
{\Theta\big(q^{-m}a_0^{-1};q\big)\Theta\big(q^{(n+1)/2}a_2;q\big)\big(q^{(n+2)/2}a_2;q\big)_\infty\big(q^{1/2};q\big)_\infty}
\\
\phantom{G_{n,m} =+}
\times {}_2\varphi_1\left(
\begin{matrix}
0,q^{(n+2)/2}a_2
\\
q^{1/2}
\end{matrix}
;q,q^{-m+1}a_0^{-1}\right),
\end{gather*}
where $\Lambda_{n,m}$ is a~periodic function of period two for~$n$ and period one for~$m$, i.e.,
\begin{gather*}
\Lambda_{n+2,m}=\Lambda_{n,m+1}=\Lambda_{n,m}.
\end{gather*}
\end{lemma}

\begin{proof}
For convenience, we put
\begin{gather*}
t=q^{-m+1}a_0^{-1},
\qquad
\alpha=q^{n/2}a_2,
\qquad
G_{n,m}=G(t,\alpha).
\end{gather*}
Then, equations~\eqref{eqn:relation_G_2} and~\eqref{eqn:relation_G_123} can be rewritten as
\begin{gather}
 tG\big(t,q^{-1}\alpha\big)+G\big(t,q^{-3/2}\alpha\big)-\alpha G\big(qt,q^{-3/2}\alpha\big)=0,
\label{eqn:relation_Ft_2}
\\
 G\big(q^2t,\alpha\big)+\big(t-\big(1+q^{1/2}\big)q^{-3/2}\alpha^{-1}\big)G(qt,\alpha) -q^{-5/2}\alpha^{-2}(t-1)G(t,\alpha)=0,
\label{eqn:relation_Ft_123}
\end{gather}
respectively.
Substituting
\begin{gather*}
G(t,\alpha)=D(t,\alpha)\sum\limits^{\infty}_{k=0}C_k(\alpha)t^k,
\end{gather*}
in equation~\eqref{eqn:relation_Ft_123}, we obtain
\begin{gather}
G(t,\alpha) =D_1(t,\alpha)\,{}_2\varphi_1\left(
\begin{matrix}
0,q^{3/2}\alpha
\\
q^{3/2}
\end{matrix}
;q,t\right) +D_2(t,\alpha)\,{}_2\varphi_1\left(
\begin{matrix}
0,q\alpha
\\
q^{1/2}
\end{matrix}
;q,t\right),
\label{eqn:m-shift_F_1}
\end{gather}
where $D_1(t,\alpha)$ and $D_2(t,\alpha)$ satisfy
\begin{gather}
 D_1(qt,\alpha)=q^{-1}\alpha^{-1}D_1(t,\alpha),
\label{eqn:m-shift_d_1}
\\
 D_2(qt,\alpha)=q^{-3/2}\alpha^{-1}D_2(t,\alpha),
\label{eqn:m-shift_d_2}
\end{gather}
respectively.
Substituting~\eqref{eqn:m-shift_F_1} in equation~\eqref{eqn:relation_Ft_2}, we can obtain the following equations:
\begin{gather}
 {}_2\varphi_1\left(
\begin{matrix}
0,q^{-1/2}\alpha
\\
q^{1/2}
\end{matrix}
;q,t\right) -{}_2\varphi_1\left(
\begin{matrix}
0,q^{-1/2}\alpha
\\
q^{1/2}
\end{matrix}
;q,qt\right) =-t\frac{D_1\big(t,q^{-1}\alpha\big)}{D_2\big(t,q^{-3/2}\alpha\big)}\, {}_2\varphi_1\left(
\begin{matrix}
0,q^{1/2}\alpha
\\
q^{3/2}
\end{matrix}
;q,t\right)\!,\!\!\!\!
\label{eqn:sol_G_1}
\\
 t\frac{D_2\big(t,q^{-1}\alpha\big)}{D_1\big(t,q^{-3/2}\alpha\big)}\, {}_2\varphi_1\left(
\begin{matrix}
0,\alpha
\\
q^{1/2}
\end{matrix}
;q,t\right) +{}_2\varphi_1\left(
\begin{matrix}
0,\alpha
\\
q^{3/2}
\end{matrix}
;q,t\right) -q^{1/2}{}_2\varphi_1\left(
\begin{matrix}
0,\alpha
\\
q^{3/2}
\end{matrix}
;q,qt\right)=0.
\label{eqn:sol_G_2}
\end{gather}
Therefore, we obtain
\begin{gather}
 D_1\big(t,q^{1/2}\alpha\big)=-\frac{1-q\alpha}{1-q^{1/2}} D_2(t,\alpha),
\label{eqn:m-shift_d_3}
\\
 D_2\big(t,q^{1/2}\alpha\big)=-\frac{1-q^{1/2}}{t} D_1(t,\alpha),
\label{eqn:m-shift_d_4}
\end{gather}
from equations~\eqref{eqn:sol_G_1} and~\eqref{eqn:sol_G_2} by using equations~\eqref{eqn:hyper_relation_1}
and~\eqref{eqn:hyper_relation_2}, respectively.
By setting
\begin{gather*}
 D_1(t,\alpha) =\frac{\Theta(\alpha
t;q)}{\big(q^{-1/2};q\big)_\infty\big(q^{3/2}\alpha;q\big)_\infty\Theta\big(q^{-1}t;q\big)\Theta(\alpha;q)}  \Lambda(t,\alpha),
\\
 D_2(t,\alpha) =\frac{q^{1/2}\Theta(q^{1/2}\alpha
t;q)}{\big(q^{1/2};q\big)_\infty(q\alpha;q)_\infty\Theta\big(q^{-1}t;q\big)\Theta\big(q^{1/2}\alpha;q\big)}  \Lambda\big(t,q^{1/2}\alpha\big),
\end{gather*}
equations~\eqref{eqn:m-shift_d_1},~\eqref{eqn:m-shift_d_2},~\eqref{eqn:m-shift_d_3} and~\eqref{eqn:m-shift_d_4} can be
reduced to
\begin{gather*}
\Lambda(t,q\alpha)=\Lambda(qt,\alpha)=\Lambda(t,\alpha).
\end{gather*}
This completes the proof.
\end{proof}

\subsubsection*{Step~4.\ Constructing the closed-form expressions of $\boldsymbol{\tau_N^{n,m}}$ ($\boldsymbol{N\geq 2}$)}

In this f\/inal step, constructing the closed-form expressions of $\tau_N^{n,m}$ $(N\geq 2)$, we obtain the
hypergeometric~$\tau$-functions for the~$q$-P$_{\rm IV}$.

Let
\begin{gather*}
\tau_N^{n,m} = (-1)^{N(N-1)/2}q^{3N(N-1)(N-n+1)/4}a_2^{-3N(N-1)/2}
\left(\prod\limits_{k=1}^{N}\big(q^{(n-2k+1)/2}a_2;q^{1/2}\big)_\infty\right)
\\
\phantom{\tau_N^{n,m} =}{}
 \times \big(q^{m}a_0;q,q\big)_{\infty}\big(q^{(n+1)/2}a_2;q^{1/2},q\big)_{\infty} \Gamma\big(q^{(2n+2m-3)/4}a_0^{1/2}a_2;q^{1/2},q^{1/2}\big)
\\
\phantom{\tau_N^{n,m} =}{}
 \times\frac{\Gamma\big(q^{(n-m+1)/4}a_0^{-1/4}a_2^{1/2};q^{1/4},q^{1/4}\big)
\Gamma\big(q^{(n-m)/4}a_0^{-1/4}a_2^{1/2};q^{1/4},q^{1/4}\big)}
{\Gamma\big({-}q^{3m-1}a_0^{3};q^3,q^3\big)\Gamma\big({-}q^{2n}a_2^4;q^2,q^2\big)} \phi_N^{n,m}.
\end{gather*}
From~\eqref{eqn:qp4_boundary_condition},~\eqref{eqn:gage_qp4_tau1} and~\eqref{eqn:gage_qp4_tau0}, we get
\begin{gather*}
\phi_N^{n,m}=0,
\qquad
N<0,
\qquad
\phi_0^{n,m}=1,
\qquad
\phi_1^{n,m}=F_{n,m}.
\end{gather*}
Moreover, it is easily verif\/ied that equation~\eqref{eqn:bilinear_2} can be rewritten as
\begin{gather}
\label{eqn:qp4_toda_bilinear}
\phi_{N+1}^{n,m}\phi_{N-1}^{n-1,m} -\phi_{N}^{n-1,m}\phi_{N}^{n,m} +\phi_{N}^{n-2,m}\phi_{N}^{n+1,m}=0.
\end{gather}
In general, equation~\eqref{eqn:qp4_toda_bilinear} admits a~solution expressed in terms of Jacobi--Trudi type determinant
\begin{gather*}
\phi^{n,m}_{N} = \det (c_{n-2i+j+1,m} )_{i,j=1,\ldots,N},
\qquad
N>0,
\end{gather*}
under the boundary conditions
\begin{gather*}
\phi^{n,m}_{N} = 0,
\qquad
N<0,
\qquad
\phi^{n,m}_0 = 1,
\qquad
\phi^{n,m}_1 = c_{n,m},
\end{gather*}
where $c_{n,m}$ is an arbitrary function.
Therefore, we obtain the following lemma:
\begin{lemma}
\label{lemma:qp4_hypergeometric_tau_1}
Under the assumptions
\begin{gather*}
a_0a_1=q,
\qquad
\tau_N^{n,m}=0,
\qquad
N<0,
\end{gather*}
the hypergeometric~$\tau$-functions for the~$q$-{\rm P}$_{\rm IV}$ are given as
\begin{gather*}
\tau_N^{n,m} = (-1)^{N(N-1)/2}q^{3N(N-1)(N-n+1)/4}a_2^{-3N(N-1)/2}
\left(\prod\limits_{k=1}^N\big(q^{(n-2k+1)/2}a_2;q^{1/2}\big)_\infty\right)
\\
\phantom{\tau_N^{n,m} =}{}
 \times \big(q^{m}a_0;q,q\big)_{\infty}\big(q^{(n+1)/2}a_2;q^{1/2},q\big)_{\infty} \Gamma\big(q^{(2n+2m-3)/4}a_0^{1/2}a_2;q^{1/2},q^{1/2}\big)
\\
\phantom{\tau_N^{n,m} =}{}
 \times\frac{\Gamma\big(q^{(n-m+1)/4}a_0^{-1/4}a_2^{1/2};q^{1/4},q^{1/4}\big)
\Gamma\big(q^{(n-m)/4}a_0^{-1/4}a_2^{1/2};q^{1/4},q^{1/4}\big)}
{\Gamma\big({-}q^{3m-1}a_0^{3};q^3,q^3\big)\Gamma\big({-}q^{2n}a_2^4;q^2,q^2\big)} \phi_N^{n,m},
\end{gather*}
where
\begin{gather*}
\phi_N^{n,m}=
\begin{vmatrix}
F_{n,m}&F_{n+1,m}&\dots&F_{n+N-1,m}
\\
F_{n-2,m}&F_{n-1,m}&\dots&F_{n+N-3,m}
\\
\vdots&\vdots&\ddots&\vdots
\\
F_{n-2N+2,m}&F_{n-2N+3,m}&\dots&F_{n-N+1,m}
\end{vmatrix}
,
\quad
\phi_0^{n,m}=1,
\quad
\phi_{-N}^{n,m}=0,
\quad
N>0.
\end{gather*}
Here, $F_{n,m}$ is given in Lemma~{\rm \ref{lemma:sol_F1_F2}}.
\end{lemma}

We also show another expression of the hypergeometric~$\tau$-functions for the~$q$-P$_{\rm IV}$.
From relation~\eqref{define:Gnm}, $\phi_N^{n,m}$ can be rewritten as
\begin{gather*}
\phi_N^{n,m}= \Theta\big(q^ma_0;q\big)^N \left(\prod\limits_{k=0}^{N-1}
\frac{\Theta\big(q^{(n+k)/2}a_2;q^{1/2}\big)}{\Theta\big({-}q^{(n+k)/2}a_2;q^{1/2}\big)}\right)  \psi_N^{n-3,m-1},
\end{gather*}
where
\begin{gather*}
\psi_N^{n,m}=
\begin{vmatrix}
G_{n,m}&G_{n+1,m}&\dots&G_{n+N-1,m}
\\
G_{n-2,m}&G_{n-1,m}&\dots&G_{n+N-3,m}
\\
\vdots&\vdots&\ddots&\vdots
\\
G_{n-2N+2,m}&G_{n-2N+3,m}&\dots&G_{n-N+1,m}
\end{vmatrix}
,
\quad
\psi_0^{n,m}=1,
\quad
\psi_{-N}^{n,m}=0,
\quad
N>0.
\end{gather*}
This gives the following lemma:
\begin{lemma}
\label{lemma:qp4_hypergeometric_tau_2}
Under the assumptions
\begin{gather*}
a_0a_1=q,
\qquad
\tau_N^{n,m}=0,
\qquad
N<0,
\end{gather*}
the hypergeometric~$\tau$-functions for the~$q$-{\rm P}$_{\rm IV}$ are given as
\begin{gather*}
\tau_N^{n,m} = (-1)^{N(N-1)/2}q^{3N(N-1)(N-n+1)/4}a_2^{-3N(N-1)/2}
\left(\prod\limits_{k=1}^N\big(q^{(n-2k+1)/2}a_2;q^{1/2}\big)_\infty\right)
\\
\phantom{\tau_N^{n,m} =}{}
 \times \big(q^{m}a_0;q,q\big)_{\infty}\big(q^{(n+1)/2}a_2;q^{1/2},q\big)_{\infty} \Gamma\big(q^{(2n+2m-3)/4}a_0^{1/2}a_2;q^{1/2},q^{1/2}\big)
\\
\phantom{\tau_N^{n,m} =}{}
 \times\frac{\Gamma\big(q^{(n-m+1)/4}a_0^{-1/4}a_2^{1/2};q^{1/4},q^{1/4}\big)
\Gamma\big(q^{(n-m)/4}a_0^{-1/4}a_2^{1/2};q^{1/4},q^{1/4}\big)}
{\Gamma\big({-}q^{3m-1}a_0^{3};q^3,q^3\big)\Gamma\big({-}q^{2n}a_2^4;q^2,q^2\big)}
\\
\phantom{\tau_N^{n,m} =}{}
 \times\Theta\big(q^ma_0;q\big)^N \left(\prod\limits_{k=0}^{N-1}
\frac{\Theta\big(q^{(n+k)/2}a_2;q^{1/2}\big)}{\Theta\big({-}q^{(n+k)/2}a_2;q^{1/2}\big)}\right)  \psi_N^{n-3,m-1},
\end{gather*}
where
\begin{gather*}
\psi_N^{n,m}=
\begin{vmatrix}
G_{n,m}&G_{n+1,m}&\dots&G_{n+N-1,m}
\\
G_{n-2,m}&G_{n-1,m}&\dots&G_{n+N-3,m}
\\
\vdots&\vdots&\ddots&\vdots
\\
G_{n-2N+2,m}&G_{n-2N+3,m}&\dots&G_{n-N+1,m}
\end{vmatrix}
,
\quad
\psi_0^{n,m}=1,
\quad
\psi_{-N}^{n,m}=0,
\quad
N>0.
\end{gather*}
Here, $G_{n,m}$ is given in Lemma~{\rm \ref{lemma:sol_G2_G123}}.
\end{lemma}

\subsection[Hypergeometric solutions of the~$q$-P$_{\rm IV}$]{Hypergeometric solutions
of the~$\boldsymbol{q}$-P$\boldsymbol{_{\rm IV}}$}

In this section, we show the hypergeometric solutions of~$q$-P$_{\rm IV}$~\eqref{eqn:qp4}.

From relations~\eqref{eqn:relation_X_f2} and~\eqref{eqn:relation_f2_tau},
the variable for~$q$-P$_{\rm IV}$~\eqref{eqn:qp4} is expressed by the~$\tau$-functions as the following:
\begin{gather*}
X_n(m,N)=q^{(m+N-1)/2}a_0a_1^{1/2} \frac{\tau_N^{n,m}\tau_{N+1}^{n+3,m+1}}{\tau_{N+1}^{n+2,m+1}\tau_N^{n+1,m}}.
\end{gather*}
Therefore, from Lemmas~\ref{lemma:qp4_hypergeometric_tau_1} and~\ref{lemma:qp4_hypergeometric_tau_2}, we obtain the
following theorems:
\begin{theorem}
\label{theorem:lattice_sol_1}
The hypergeometric solutions of~$q$-{\rm P}$_{\rm IV}$~\eqref{eqn:qp4} with
\begin{gather}
\label{eqn:condition_theorem}
a_0a_1=q,
\qquad
N\geq 0,
\end{gather}
is given~by
\begin{gather*}
X_n(m,N)=-q^{(-2N-m+1)/2}a_0^{-1/2} \frac{\phi_N^{n,m}\phi_{N+1}^{n+3,m+1}}{\phi_{N+1}^{n+2,m+1}\phi_N^{n+1,m}},
\end{gather*}
where
\begin{gather*}
\phi_N^{n,m}=
\begin{vmatrix}
F_{n,m}&F_{n+1,m}&\dots&F_{n+N-1,m}
\\
F_{n-2,m}&F_{n-1,m}&\dots&F_{n+N-3,m}
\\
\vdots&\vdots&\ddots&\vdots
\\
F_{n-2N+2,m}&F_{n-2N+3,m}&\dots&F_{n-N+1,m}
\end{vmatrix}
,
\qquad
\phi_0^{n,m}=1.
\end{gather*}
Here, $F_{n,m}$ is given in Lemma~{\rm \ref{lemma:sol_F1_F2}}.
\end{theorem}
\begin{theorem}
\label{theorem:lattice_sol_2}
The hypergeometric solutions of~$q$-{\rm P}$_{\rm IV}$~\eqref{eqn:qp4} with the condition~\eqref{eqn:condition_theorem}
is given~by
\begin{gather*}
X_n(m,N)=q^{(-2N-m+1)/2}a_0^{-1/2} \frac{\psi_N^{n-3,m-1}\psi_{N+1}^{n,m}}{\psi_{N+1}^{n-1,m}\psi_N^{n-2,m-1}},
\end{gather*}
where
\begin{gather*}
\psi_N^{n,m}=
\begin{vmatrix}
G_{n,m}&G_{n+1,m}&\dots&G_{n+N-1,m}
\\
G_{n-2,m}&G_{n-1,m}&\dots&G_{n+N-3,m}
\\
\vdots&\vdots&\ddots&\vdots
\\
G_{n-2N+2,m}&G_{n-2N+3,m}&\dots&G_{n-N+1,m}
\end{vmatrix}
,
\qquad
\psi_0^{n,m}=1.
\end{gather*}
Here, $G_{n,m}$ is given in Lemma~{\rm \ref{lemma:sol_G2_G123}}.
\end{theorem}

\section[Hypergeometric solutions of the~$q$-P$_{\rm IV}$ (II)]{Hypergeometric solutions
of the~$\boldsymbol{q}$-P$\boldsymbol{_{\rm IV}}$ (II)}
\label{section:hyper_sol_qP4_2}

In this section, we show that~$q$-P$_{\rm IV}$~\eqref{eqn:qp4} also has the hypergeometric solutions expressed~by
bilateral basic hypergeometric series.

First, we recall the def\/initions of orthogonal polynomials.
\begin{definition}
A~polynomial sequence $(P_n(t))_{n=0}^\infty$ which satisf\/ies the following conditions is called an orthogonal
polynomial sequence over the f\/ield ${\cal K}$, and each term $P_n(t)$ is called an orthogonal polynomial over the f\/ield
${\cal K}$.
\begin{enumerate}\itemsep=0pt
\item[(i)] deg$(P_n(t))=n$;
\item[(ii)] there exists a~linear functional ${\cal L}:{\cal K}(t)\to {\cal K}$ which holds the orthogonal condition:
\begin{gather*}
{\cal L}[t^kP_n(t)]=h_n\delta_{n,k},
\qquad
n\geq k,
\end{gather*}
where $\delta_{n,k}$ is Kronecker's symbol.
Here, $h_n$ and $\mu_n={\cal L}[t^n]$ are called a~normalization factor and a~moment, respectively.
\end{enumerate}
\end{definition}
\begin{definition}
An orthogonal polynomial sequence whose leading coef\/f\/icient is $1$ is called a~monic orthogonal polynomial sequence
(MOPS).
Let $(P_n(t))_{n=0}^\infty$ be a~MOPS.
Then, polynomial $P_n(t)$ and its normalization factor $h_n$ are expressed by the moment $\mu_n$ as the following:
\begin{gather}
 P_n =\frac{1}{\Phi_n}
\begin{vmatrix}
\mu_0&\mu_1&\dots&\mu_{n-1}&\mu_n
\\
\mu_1&\mu_2&\dots&\mu_n&\mu_{n+1}
\\
\vdots&\vdots&\ddots&\vdots&\vdots
\\
\mu_{n-1}&\mu_n&\dots&\mu_{2n-2}&\mu_{2n-1}
\\
1&t&\dots&t^{n-1}&t^n
\end{vmatrix}
,
\qquad\!
P_0=1,
\qquad\!
 h_n=\frac{\Phi_{n+1}}{\Phi_n},
\qquad\!
h_0=\mu_0,\!\!\!
\label{eqn:appendix_hn_determinant}
\end{gather}
where $\Phi_n$ is the Hankel determinant given~by
\begin{gather}
\label{eqn:appendix_Phi_determinant}
\Phi_n=
\begin{vmatrix}
\mu_0&\mu_1&\dots&\mu_{n-1}
\\
\mu_1&\mu_2&\dots&\mu_n
\\
\vdots&\vdots&\ddots&\vdots
\\
\mu_{n-1}&\mu_n&\dots&\mu_{2n-2}
\end{vmatrix}
.
\end{gather}
\end{definition}

We assume that $(P_n)_{n=0}^\infty=(P_n(t))_{n=0}^\infty$ and $(\hat{P}_n)_{n=0}^\infty=(\hat{P}_n(t))_{n=0}^\infty$ are
MOPSs which satisfy the following orthogonal conditions:
\begin{gather*}
 {\cal L}\big[t^kP_n(t)\big]=h_n\delta_{n,k},
\quad
n\geq k,  \qquad
 \hat{\cal L}\big[t^k\hat{P}_n(t)\big]=\hat{h}_n\delta_{n,k},
\quad
n\geq k,
\end{gather*}
respectively.
In addition, we put the case that $P_n$ and $\hat{P}_n$ are related by the Christof\/fel transformation (or Geronimus
transformation), that is, the linear functionals satisfy the following relation for an arbitrary function~$f(t)$:
\begin{gather*}
{\cal L}[f(t)]=\hat{\cal L}\left[\frac{f(t)}{t-c_0}+\delta(t-c_0)\right],
\end{gather*}
where $\delta(x)$ is the Dirac delta function and $c_0\in\mathbb{C}$ is an additional parameter.
For these MOPSs, the following relations
hold~\cite{IsmailMEH2009:MR2542683,UvarovVB1969:MR0262764,ZhedanovA1997:MR1482157}:
\begin{gather}
 (t-c_0)\hat{P}_n=P_{n+1}+\frac{\hat{h}_n}{h_n} P_n,
\label{eqn:P_n:1}
\\
 P_n=\hat{P}_n+\frac{h_n}{\hat{h}_{n-1}} \hat{P}_{n-1}.
\label{eqn:P_n:2}
\end{gather}
Eliminating $P_n$ from equation~\eqref{eqn:P_n:1} by using equation~\eqref{eqn:P_n:2}, we obtain the following
three-term recurrence relation:
\begin{gather}
\label{eqn:3term_hatPn_1}
t\hat{P}_n =\hat{P}_{n+1}+\left(\frac{h_{n+1}}{\hat{h}_n}+\frac{\hat{h}_n}{h_n}+c_0\right)\hat{P}_n
+\frac{\hat{h}_n}{\hat{h}_{n-1}} \hat{P}_{n-1}.
\end{gather}
Let
\begin{gather*}
\hat{P}_n(t)=\frac{\tilde{h}_n(c_1t;p)}{c_1^n},
\qquad
c_1>0,
\end{gather*}
where $\tilde{h}_n(t;q)$ is the discrete~$q$-Hermite II polynomial:
\begin{gather*}
\tilde{h}_n(t;q)=t^n {}_2\varphi_1\left(
\begin{matrix}
q^{-n},q^{-n+1}
\\
0
\end{matrix}
;q^2,-\frac{q^2}{t^2}\right).
\end{gather*}
Then, the linear functional, the normalization factor and the three-term recurrence relation for~$\hat{P}_n$ are given
by
\begin{gather}
 \hat{\cal L}[f(t)]=\int^\infty_{-\infty}\frac{f(t)}{\big({-}c_1^2t^2;p^2\big)_\infty} {\rm d}_pt,
\nonumber
\\
 \hat{h}_n
=\frac{2}{p^{n^2}c_1^{2n}} \frac{(p;p)_n\big(p^2;p^2\big)_\infty\Theta\big({-}pc_1^2;p^2\big)}{\big(p^3;p^2\big)_\infty\Theta\big({-}c_1^2;p^2\big)},
\label{eqn:appendix_hat_hn}
\\
 t\hat{P}_n=\hat{P}_{n+1}+p^{-2n+1}\big(1-p^n\big)c_1^{-2} \hat{P}_{n-1},
\label{eqn:appendix_three_term}
\end{gather}
respectively.
We note that these properties of~$q$-Hermite II polynomials are given in~\cite{book_KLS2010:MR2656096}.
We here impose the condition $c_0\neq p^a$ for all $a\in\mathbb{Z}$ since the linear functional for $P_n$ is given~by
\begin{gather*}
{\cal L}[f(t)]=\int^\infty_{-\infty}\frac{f(t)}{(t-c_0)\big({-}c_1^2t^2;p^2\big)_\infty} {\rm d}_pt.
\end{gather*}
In addition, the moment $\mu_n$ can be obtained~by
\begin{gather}
\mu_n=-\frac{1-p}{c_0^2} \sum\limits_{k=-\infty}^\infty
\frac{\big(1-(-1)^n\big)p^k+\big(1+(-1)^n\big)c_0}{\big(1-p^{2k}c_0^{-2}\big)\big({-}c_1^2p^{2k};p^2\big)_\infty} p^{k(n+1)}
\nonumber
\\
\hphantom{\mu_n}{}
 =\frac{2(1-p)}{\big(1-c_0^{2}\big)\big({-}c_1^2;p^2\big)_\infty} \sum\limits_{k=-\infty}^\infty
\frac{\big({-}c_1^2,c_0^{-2};p^2\big)_k}{\big(p^2c_0^{-2};p^2\big)_k}
\left(\frac{1-(-1)^n}{2} p^k+\frac{1+(-1)^n}{2} c_0\right)p^{k(n+1)}
\nonumber
\\
\hphantom{\mu_n}{}
 =
\begin{cases}
\dfrac{2(1-p)}{\big(1-c_0^{2}\big)\big({-}c_1^2;p^2\big)_\infty}\, {}_2\psi_2\left(
\begin{matrix}
-c_1^2,c_0^{-2}
\\
0,p^2c_0^{-2}
\end{matrix}
;p^2,p^{2k+1}\right), & n=2k-1,
\vspace{1mm}\\
\dfrac{2c_0(1-p)}{\big(1-c_0^2\big)\big({-}c_1^2;p^2\big)_\infty}\, {}_2\psi_2\left(
\begin{matrix}
-c_1^2,c_0^{-2}
\\
0,p^2c_0^{-2}
\end{matrix}
;p^2,p^{2k+1}\right), & n=2k.
\end{cases}
\label{eqn:appendix_mu}
\end{gather}
Comparing the coef\/f\/icients of equations~\eqref{eqn:3term_hatPn_1} and~\eqref{eqn:appendix_three_term}, we obtain the
following equations:
\begin{gather}
 \frac{h_{n+1}}{\hat{h}_n}+\frac{\hat{h}_n}{h_n}+c_0=0,
\label{eqn:appendix_h_1}
\\
 \frac{\hat{h}_n}{\hat{h}_{n-1}}=p^{-2n+1}\big(1-p^n\big)c_1^{-2}.
\label{eqn:appendix_h_2}
\end{gather}
From equations~\eqref{eqn:appendix_h_1} and~\eqref{eqn:appendix_h_2}, we obtain
\begin{gather}
\label{eqn:appendix_h_3}
\frac{\hat{h}_n}{h_{n+1}}=-\frac{h_n}{p^{-2n+1}(1-p^n)c_1^{-2}\hat{h}_{n-1}+c_0h_n}.
\end{gather}
By setting
\begin{gather}
\label{eqn:appendix_X}
X_n={\rm i}\frac{1-p^{n+1}}{p^nc_1} \frac{\hat{h}_n}{h_{n+1}},
\end{gather}
equation~\eqref{eqn:appendix_h_3} can be rewritten as the following discrete Riccati equation:
\begin{gather}
\label{eqn:appendix_riccati}
X_n=\frac{1-p^{n+1}}{X_{n-1}+{\rm i} p^nc_1c_0}.
\end{gather}
Since in the case of
\begin{gather*}
a_0^{1/2}a_1^{1/2}=q^{1/2},
\qquad
a_2=1,
\qquad
N=-1,
\end{gather*}
$q$-P$_{\rm IV}$~\eqref{eqn:qp4} admits the reduction to
\begin{gather*}
X_n=\frac{1-q^{(n+1)/2}}{X_{n-1}+q^{(n-m+1)/2}a_0^{-1/2}},
\end{gather*}
which is equivalent to equation~\eqref{eqn:appendix_riccati} with the following correspondence:
\begin{gather*}
q^{m/2}a_0^{1/2}=-{\rm i} q^{1/2}c_0^{-1}c_1^{-1},
\qquad
q^{1/2}=p,
\end{gather*}
\eqref{eqn:appendix_hn_determinant},~\eqref{eqn:appendix_Phi_determinant},~\eqref{eqn:appendix_hat_hn},~\eqref{eqn:appendix_mu}
and~\eqref{eqn:appendix_X} give the hypergeometric solutions of~$q$-P$_{\rm IV}$~\eqref{eqn:qp4}.
Therefore, we f\/inally obtain the following theorem:
\begin{theorem}
\label{theorem:molecule_sol}
In the case of
\begin{gather*}
a_0^{1/2}a_1^{1/2}=q^{1/2},
\qquad
a_2=1,
\qquad
N=-1,
\qquad
n\geq0,
\end{gather*}
$q$-{\rm P}$_{\rm IV}$~\eqref{eqn:qp4} with
\begin{gather*}
q^{m/2}a_0^{1/2}=-{\rm i} q^{1/2} c_0^{-1}c_1^{-1},
\end{gather*}
admits the following hypergeometric solution:
\begin{gather*}
X_n= \frac{2{\rm i}\big(1-q^{(n+1)/2}\big)\big(q^{1/2};q^{1/2}\big)_n(q;q)_\infty\Theta\big({-}q^{1/2}c_1^2;q\big)}
{q^{n(n+1)/2}c_1^{2n+1}\big(q^{3/2};q\big)_\infty\Theta\big({-}c_1^2;q\big)}  \frac{\Phi_{n+1}}{\Phi_{n+2}},
\end{gather*}
where
\begin{gather*}
 \Phi_n=
\begin{vmatrix}
\mu_0&\mu_1&\dots&\mu_{n-1}
\\
\mu_1&\mu_2&\dots&\mu_n
\\
\vdots&\vdots&\ddots&\vdots
\\
\mu_{n-1}&\mu_n&\dots&\mu_{2n-2}
\end{vmatrix}
,
\\
\mu_n =
\begin{cases}
\dfrac{2\big(1-q^{1/2}\big)}{\big(1-c_0^{2}\big)\big({-}c_1^2;q\big)_\infty}\, {}_2\psi_2\left(
\begin{matrix}
-c_1^2,c_0^{-2}
\\
0,qc_0^{-2}
\end{matrix}
;q,q^{(2k+1)/2}\right), & n=2k-1,
\vspace{1mm}\\
\dfrac{2c_0\big(1-q^{1/2}\big)}{\big(1-c_0^2\big)\big({-}c_1^2;q\big)_\infty}\, {}_2\psi_2\left(
\begin{matrix}
-c_1^2,c_0^{-2}
\\
0,qc_0^{-2}
\end{matrix}
;q,q^{(2k+1)/2}\right), & n=2k.
\end{cases}
\end{gather*}
Here, $c_0\neq q^{a/2}$ for all $a\in\mathbb{Z}$.
\end{theorem}

\section{Concluding remarks}
\label{section:concluding}

In this paper, we have constructed the hypergeometric solutions of~$q$-P$_{\rm IV}$~\eqref{eqn:qp4} via the construction
of the hypergeometric~$\tau$-functions and the theory of orthogonal polynomials.
We showed that the hypergeometric solutions of the~$q$-P$_{\rm IV}$ can be expressed by the three expressions.
We note that the hypergeometric solutions of Painlev\'e systems expressed by the determinants whose sizes do not depend
on the independent variable are called the lattice type solutions, while those expressed by the determinants whose sizes
depend on the independent variable are called molecule type solutions.
Thus, the hypergeometric solutions given in Theorems~\ref{theorem:lattice_sol_1} and~\ref{theorem:lattice_sol_2} are
lattice type solutions whereas those given in Theorem~\ref{theorem:molecule_sol} are molecule type solutions.

Before closing, we mention the bilateral type hypergeometric solutions here.
It is well known that the coalescence cascade of hypergeometric functions, from the Gauss's hypergeometric function to
the Airy function, corresponds to the diagram of degeneration of the Painlev\'e equations, from the Painlev\'e~VI
equation to the Painlev\'e~II equation, in the sense of the hypergeometric solutions~\cite{book_IKKSY1991:MR1118604}:
\begin{gather*}
\begin{matrix}
{\rm P}_{\rm VI}&\to&{\rm P}_{\rm V}&\to&{\rm P}_{\rm III}
\\
\text{Gauss}&&\text{Kummer}&&\text{Bessel}
\\
&&\downarrow&&\downarrow
\\
&&{\rm P}_{\rm IV}&\to&{\rm P}_{\rm II}&
\\
&&\text{Weber}&&\text{Airy}
\end{matrix}
\end{gather*}
Similarly, the relations between basic hypergeometric series and~$q$-Painlev\'e equations are also
investigated~\cite{KMNOY2005:MR2153786,KN:inpress}.
However, the hypergeometric solutions described by bilateral basic hypergeometric series have not been considered.
It might be an interesting future problem to make a~list of the bilateral basic hypergeometric series that appear as the
solutions of the~$q$-Painlev\'e equations.

\subsubsection*{Acknowledgments}

The author would like to thank Professors K.~Kajiwara, S.~Kakei, H.~Miki, M.~Noumi, and S.~Tsujimoto for the useful
comments.
He also appreciates the valuable comments from the referees which have improved the quality of this paper.
This work has been supported by JSPS Grant-in-Aid for Scientif\/ic Research No.~22$\cdot$4366 and the Australian Research
Council grant DP130100967.

\pdfbookmark[1]{References}{ref}
\LastPageEnding


\begin{thebibliography}{99}
\footnotesize \itemsep=-1pt

\bibitem{book_GR2004:MR2128719}
Gasper G., Rahman M., Basic hypergeometric series, \href{http://dx.doi.org/10.1017/CBO9780511526251}{\textit{Encyclopedia of
  Mathematics and its Applications}}, Vol.~96, 2nd ed., Cambridge University
  Press, Cambridge, 2004.

\bibitem{HK2007:MR2392886}
Hamamoto T., Kajiwara K., Hypergeometric solutions to the {$q$}-{P}ainlev\'e
  equation of type {$A_4^{(1)}$}, \href{http://dx.doi.org/10.1088/1751-8113/40/42/S01}{\textit{J.~Phys.~A: Math. Gen.}} \textbf{40}
  (2007), 12509--12524, \href{http://arxiv.org/abs/nlin.SI/0701001}{nlin.SI/0701001}.

\bibitem{HKW2006:MR2264726}
Hamamoto T., Kajiwara K., Witte N.S., Hypergeometric solutions to the
  {$q$}-{P}ainlev\'e equation of type {$(A_1+A'_1)^{(1)}$}, \href{http://dx.doi.org/10.1155/IMRN/2006/84619}{\textit{Int. Math.
  Res. Not.}} \textbf{2006} (2006), 84619, 26~pages, \href{http://arxiv.org/abs/nlin.SI/0607065}{nlin.SI/0607065}.

\bibitem{IsmailMEH2009:MR2542683}
Ismail M.E.H., Classical and quantum orthogonal polynomials in one variable,
  \href{http://dx.doi.org/10.1017/CBO9781107325982}{\textit{Encyclopedia of Mathematics and its Applications}}, Vol.~98, Cambridge
  University Press, Cambridge, 2005.

\bibitem{book_IKKSY1991:MR1118604}
Iwasaki K., Kimura H., Shimomura S., Yoshida M., From {G}auss to {P}ainlev\'e.
  A modern theory of special functions, \href{http://dx.doi.org/10.1007/978-3-322-90163-7}{\textit{Aspects of Mathematics}}, Vol.~E16, Friedr. Vieweg \& Sohn, Braunschweig, 1991.

\bibitem{JM1981:MR625446}
Jimbo M., Miwa T., Monodromy preserving deformation of linear ordinary
  dif\/ferential equations with rational coef\/f\/icients.~{II}, \href{http://dx.doi.org/10.1016/0167-2789(81)90021-X}{\textit{Phys.~D}}
  \textbf{2} (1981), 407--448.

\bibitem{JM1981:MR636469}
Jimbo M., Miwa T., Monodromy preserving deformation of linear ordinary
  dif\/ferential equations with rational coef\/f\/icients.~{III}, \href{http://dx.doi.org/10.1016/0167-2789(81)90003-8}{\textit{Phys.~D}}
  \textbf{4} (1981), 26--46.

\bibitem{JMU1981:MR630674}
Jimbo M., Miwa T., Ueno K., Monodromy preserving deformation of linear ordinary
  dif\/ferential equations with rational coef\/f\/icients. {I}.~{G}eneral theory and
  {$\tau$}-function, \href{http://dx.doi.org/10.1016/0167-2789(81)90013-0}{\textit{Phys.~D}} \textbf{2} (1981), 306--352.

\bibitem{JKM2006:MR2297948}
Joshi N., Kajiwara K., Mazzocco M., Generating function associated with the
  {H}ankel determinant formula for the solutions of the {P}ainlev\'e {IV}
  equation, \href{http://dx.doi.org/10.1619/fesi.49.451}{\textit{Funkcial. Ekvac.}} \textbf{49} (2006), 451--468,
  \href{http://arxiv.org/abs/nlin.SI/0512041}{nlin.SI/0512041}.

\bibitem{KK2003:MR1952868}
Kajiwara K., Kimura K., On a {$q$}-dif\/ference {P}ainlev\'e {III} equation.
  {I}.~{D}erivation, symmetry and {R}iccati type solutions,
  \href{http://dx.doi.org/10.2991/jnmp.2003.10.1.7}{\textit{J.~Nonlinear Math. Phys.}} \textbf{10} (2003), 86--102,
  \href{http://arxiv.org/abs/nlin.SI/0205019}{nlin.SI/0205019}.

\bibitem{KM1999:MR1694666}
Kajiwara K., Masuda T., A generalization of determinant formulae for the
  solutions of {P}ainlev\'e {II} and {XXXIV} equations, \href{http://dx.doi.org/10.1088/0305-4470/32/20/309}{\textit{J.~Phys.~A:
  Math. Gen.}} \textbf{32} (1999), 3763--3778, \href{http://arxiv.org/abs/solv-int/9903014}{solv-int/9903014}.

\bibitem{KMNOY2003:MR1984002}
Kajiwara K., Masuda T., Noumi M., Ohta Y., Yamada Y., {${}_{10}E_9$} solution
  to the elliptic {P}ainlev\'e equation, \href{http://dx.doi.org/10.1088/0305-4470/36/17/102}{\textit{J.~Phys.~A: Math. Gen.}}
  \textbf{36} (2003), L263--L272, \href{http://arxiv.org/abs/nlin.SI/0303032}{nlin.SI/0303032}.

\bibitem{KMNOY2004:MR2077840}
Kajiwara K., Masuda T., Noumi M., Ohta Y., Yamada Y., Hypergeometric solutions
  to the {$q$}-{P}ainlev\'e equations, \href{http://dx.doi.org/10.1155/S1073792804140919}{\textit{Int. Math. Res. Not.}}
  \textbf{2004} (2004), 2497--2521, \href{http://arxiv.org/abs/nlin.SI/0403036}{nlin.SI/0403036}.

\bibitem{KMNOY2005:MR2153786}
Kajiwara K., Masuda T., Noumi M., Ohta Y., Yamada Y., Construction of
  hypergeometric solutions to the {$q$}-{P}ainlev\'e equations, \href{http://dx.doi.org/10.1155/IMRN.2005.1439}{\textit{Int.
  Math. Res. Not.}} \textbf{2005} (2005), 1441--1463, \href{http://arxiv.org/abs/nlin.SI/0501051}{nlin.SI/0501051}.

\bibitem{KMNOY2006:MR2353465}
Kajiwara K., Masuda T., Noumi M., Ohta Y., Yamada Y., Point conf\/igurations,
  {C}remona transformations and the elliptic dif\/ference {P}ainlev\'e equation,
  in Th\'eories asymptotiques et \'equations de {P}ainlev\'e, \textit{S\'emin.
  Congr.}, Vol.~14, Soc. Math. France, Paris, 2006, 169--198,
  \href{http://arxiv.org/abs/nlin.SI/0411003}{nlin.SI/0411003}.

\bibitem{KN:inpress}
Kajiwara K., Nakazono N., Hypergeometric solutions to the symmetric
  {$q$}-{P}ainlev\'e equations, \href{http://dx.doi.org/10.1093/imrn/rnt237}{\textit{Int. Math. Res. Not.}}, {t}o appear,
  \href{http://arxiv.org/abs/1304.0858}{arXiv:1304.0858}.

\bibitem{KNT2011:MR2773334}
Kajiwara K., Nakazono N., Tsuda T., Projective reduction of the discrete
  {P}ainlev\'e system of type {$(A_2+A_1)^{(1)}$}, \href{http://dx.doi.org/10.1093/imrn/rnq089}{\textit{Int. Math. Res.
  Not.}}  (2011), 930--966, \href{http://arxiv.org/abs/0910.4439}{arXiv:0910.4439}.

\bibitem{KNY2001:MR1876614}
Kajiwara K., Noumi M., Yamada Y., A study on the fourth {$q$}-{P}ainlev\'e
  equation, \href{http://dx.doi.org/10.1088/0305-4470/34/41/312}{\textit{J.~Phys.~A: Math. Gen.}} \textbf{34} (2001), 8563--8581,
  \href{http://arxiv.org/abs/nlin.SI/0012063}{nlin.SI/0012063}.

\bibitem{KO1998:MR1629434}
Kajiwara K., Ohta Y., Determinant structure of the rational solutions for the
  {P}ainlev\'e {IV} equation, \href{http://dx.doi.org/10.1088/0305-4470/31/10/017}{\textit{J.~Phys.~A: Math. Gen.}} \textbf{31}
  (1998), 2431--2446, \href{http://arxiv.org/abs/solv-int/9709011}{solv-int/9709011}.

\bibitem{KOS1995:MR1341981}
Kajiwara K., Ohta Y., Satsuma J., Casorati determinant solutions for the
  discrete {P}ainlev\'e~{III} equation, \href{http://dx.doi.org/10.1063/1.531353}{\textit{J.~Math. Phys.}} \textbf{36}
  (1995), 4162--4174, \href{http://arxiv.org/abs/solv-int/9412004}{solv-int/9412004}.

\bibitem{KOSGR1994:MR1267458}
Kajiwara K., Ohta Y., Satsuma J., Grammaticos B., Ramani A., Casorati
  determinant solutions for the discrete {P}ainlev\'e-{II} equation,
  \href{http://dx.doi.org/10.1088/0305-4470/27/3/030}{\textit{J.~Phys.~A: Math. Gen.}} \textbf{27} (1994), 915--922,
  \href{http://arxiv.org/abs/solv-int/9310002}{solv-int/9310002}.

\bibitem{book_KLS2010:MR2656096}
Koekoek R., Lesky P.A., Swarttouw R.F., Hypergeometric orthogonal polynomials
  and their {$q$}-analogues, \href{http://dx.doi.org/10.1007/978-3-642-05014-5}{\textit{Springer Monographs in Mathematics}},
  Springer-Verlag, Berlin, 2010.

\bibitem{MasudaT2009:MR2506177}
Masuda T., Hypergeometric {$\tau$}-functions of the {$q$}-{P}ainlev\'e system
  of type {$E_7^{(1)}$}, \href{http://dx.doi.org/10.3842/SIGMA.2009.035}{\textit{SIGMA}} \textbf{5} (2009), 035, 30~pages,
  \href{http://arxiv.org/abs/0903.4102}{arXiv:0903.4102}.

\bibitem{MasudaT2011:MR2765599}
Masuda T., Hypergeometric {$\tau$}-functions of the {$q$}-{P}ainlev\'e system
  of type {$E^{(1)}_8$}, \href{http://dx.doi.org/10.1007/s11139-010-9262-1}{\textit{Ramanujan~J.}} \textbf{24} (2011), 1--31.

\bibitem{book_MJD2000:MR1736222}
Miwa T., Jimbo M., Date E., Solitons. Dif\/ferential equations, symmetries and
  inf\/inite-dimensional algebras, \textit{Cambridge Tracts in Mathematics}, Vol.~135, Cambridge University Press, Cambridge, 2000.

\bibitem{NakazonoN2010:MR2769931}
Nakazono N., Hypergeometric {$\tau$} functions of the {$q$}-{P}ainlev\'e
  systems of type {$(A_2+A_1)^{(1)}$}, \href{http://dx.doi.org/10.3842/SIGMA.2010.084}{\textit{SIGMA}} \textbf{6} (2010), 084,
  16~pages, \href{http://arxiv.org/abs/1008.2595}{arXiv:1008.2595}.

\bibitem{book_NoumiM2004:MR2044201}
Noumi M., Painlev\'e equations through symmetry, \textit{Translations of
  Mathematical Monographs}, Vol.~223, Amer. Math. Soc., Providence, RI, 2004.

\bibitem{ON1992:MR1200649}
Ohta Y., Nakamura A., Similarity {KP} equation and various dif\/ferent
  representations of its solutions, \href{http://dx.doi.org/10.1143/JPSJ.61.4295}{\textit{J.~Phys. Soc. Japan}} \textbf{61}
  (1992), 4295--4313.

\bibitem{OkamotoK1986:MR854008}
Okamoto K., Studies on the {P}ainlev\'e equations. {III}.~{S}econd and fourth
  {P}ainlev\'e equations, {$P_{{\rm II}}$} and {$P_{{\rm IV}}$}, \href{http://dx.doi.org/10.1007/BF01458459}{\textit{Math.
  Ann.}} \textbf{275} (1986), 221--255.

\bibitem{OkamotoK1987:MR916698}
Okamoto K., Studies on the {P}ainlev\'e equations. {I}.~{S}ixth {P}ainlev\'e
  equation {$P_{{\rm VI}}$}, \href{http://dx.doi.org/10.1007/BF01762370}{\textit{Ann. Mat. Pura Appl.}} \textbf{146} (1987),
  337--381.

\bibitem{OkamotoK1987:MR914314}
Okamoto K., Studies on the {P}ainlev\'e equations. {II}. {F}ifth {P}ainlev\'e
  equation {$P_{\rm V}$}, \textit{Japan.~J. Math.~(N.S.)} \textbf{13} (1987),
  47--76.

\bibitem{OkamotoK1987:MR927186}
Okamoto K., Studies on the {P}ainlev\'e equations. {IV}.~{T}hird {P}ainlev\'e
  equation {$P_{{\rm III}}$}, \textit{Funkcial. Ekvac.} \textbf{30} (1987),
  305--332.

\bibitem{QRT1988:MR924318}
Quispel G.R.W., Roberts J.A.G., Thompson C.J., Integrable mappings and soliton
  equations, \href{http://dx.doi.org/10.1016/0375-9601(88)90803-1}{\textit{Phys. Lett.~A}} \textbf{126} (1988), 419--421.

\bibitem{QRT1989:MR982386}
Quispel G.R.W., Roberts J.A.G., Thompson C.J., Integrable mappings and soliton
  equations.~{II}, \href{http://dx.doi.org/10.1016/0167-2789(89)90233-9}{\textit{Phys.~D}} \textbf{34} (1989), 183--192.

\bibitem{RG1996:MR1399286}
Ramani A., Grammaticos B., Discrete {P}ainlev\'e equations: coalescences,
  limits and degeneracies, \href{http://dx.doi.org/10.1016/0378-4371(95)00439-4}{\textit{Phys.~A}} \textbf{228} (1996), 160--171,
  \href{http://arxiv.org/abs/solv-int/9510011}{solv-int/9510011}.

\bibitem{SakaiH1998:MR1632613}
Sakai H., Casorati determinant solutions for the {$q$}-dif\/ference sixth
  {P}ainlev\'e equation, \href{http://dx.doi.org/10.1088/0951-7715/11/4/004}{\textit{Nonlinearity}} \textbf{11} (1998), 823--833.

\bibitem{SakaiH2001:MR1882403}
Sakai H., Rational surfaces associated with af\/f\/ine root systems and geometry of
  the {P}ainlev\'e equations, \href{http://dx.doi.org/10.1007/s002200100446}{\textit{Comm. Math. Phys.}} \textbf{220} (2001),
  165--229.

\bibitem{TGCR2004:MR2058894}
Tamizhmani K.M., Grammaticos B., Carstea A.S., Ramani A., The {$q$}-discrete
  {P}ainlev\'e~{IV} equations and their properties, \href{http://dx.doi.org/10.1070/RD2004v009n01ABEH000260}{\textit{Regul. Chaotic
  Dyn.}} \textbf{9} (2004), 13--20.

\bibitem{TsudaT2006:MR2207047}
Tsuda T., Tau functions of {$q$}-{P}ainlev\'e~{III} and~{IV} equations,
  \href{http://dx.doi.org/10.1007/s11005-005-0037-3}{\textit{Lett. Math. Phys.}} \textbf{75} (2006), 39--47.

\bibitem{UvarovVB1969:MR0262764}
Uvarov V.B., The connection between systems of polynomials that are orthogonal
  with respect to dif\/ferent distribution functions, \href{http://dx.doi.org/10.1016/0041-5553(69)90124-4}{\textit{USSR Comput. Math.
  Math. Phys.}} \textbf{9} (1969), no.~6, 25--36.

\bibitem{ZhedanovA1997:MR1482157}
Zhedanov A., Rational spectral transformations and orthogonal polynomials,
  \href{http://dx.doi.org/10.1016/S0377-0427(97)00130-1}{\textit{J.~Comput. Appl. Math.}} \textbf{85} (1997), 67--86.

\end{thebibliography}
\end{document}